\documentclass[journal=jpccck,manuscript=article,layout=traditional]{achemso}
\usepackage[latin9]{inputenc}
\usepackage{amsmath}
\usepackage{xcolor}
\usepackage{graphicx}

\makeatletter

\title{Light-induced nonadiabatic dissipative quantum dynamics of the $\mathrm{Na_{2}}$
molecule}

\author{Patrick Barron}

\affiliation{Department of Physics and Astronomy, Vanderbilt University, Nashville,
TN 37235, USA}

\author{Kriszti\'an Szab\'o }

\affiliation{Department of Theoretical Physics, Doctoral School of Physics, University
of Debrecen, P.O. Box 400, H-4002 Debrecen, Hungary}

\author{G\'abor J. Hal\'asz}

\affiliation{Department of Information Technology, University of Debrecen, P.O.
Box 400, H-4002 Debrecen, Hungary}

\author{K\'alm\'an Varga }

\affiliation{Department of Physics and Astronomy, Vanderbilt University, Nashville,
TN 37235, USA}

\author{\'Agnes Vib\'ok}

\affiliation{Department of Theoretical Physics, Doctoral School of Physics,University
of Debrecen, P.O. Box 400, H-4002 Debrecen, Hungary\\
ELI-ALPS, ELI-HU Non-Profit Ltd, H-6720 Szeged, Dugonics t\'er 13, Hungary}

\email{vibok@phys.unideb.hu}

\makeatother

\begin{document}
\newcommand{\revvk[1]{{\color{red} ##1}}}
\begin{abstract}
Strong light-matter coupling between molecules and optical or plasmonic
cavity modes have emerged as a promising platform for advancing photonics,
materials science, and chemistry. However, optical cavities and plasmonic
resonators in particular are inherently lossy systems characterized
by finite photon lifetimes. Accurate theoretical descriptions of molecular
dynamics under strong coupling therefore, require a proper treatment
of cavity losses.

In this work, we compare three theoretical approaches for modeling
dissipative molecule-cavity dynamics within a realistic parameter
regime: the Lindblad master equation, the stochastic Schr\"odinger equation,
and the non-Hermitian Schr\"odinger equation. As an example, we consider
the two lowest energy state of $\mathrm{Na_{2}}$ molecule coupled
to a cavity mode and analyze the time evolution of the excited-state
population and the mean photon number. Our results demonstrate that
the stochastic Schr\"odinger equation provides an accurate and computationally
efficient alternative to the Lindblad master equation, while the non-Hermitian
Schr\"odinger approach is found to be applicable only within a limited
range of conditions.

Furthermore, we show that inclusion of molecular rotation leads to
rotational-vibrational-photonic coupling and gives rise to pronounced
nonadiabatic dynamics through light-induced conical
intersections. These findings highlight the importance of both dissipation
and rotational degrees of freedom for a realistic description of molecular
dynamics in strongly coupled molecule-cavity systems.
\end{abstract}

\section*{I. Introduction}

Molecular cavity quantum electrodynamics explores the interaction
between molecules and confined electromagnetic field modes, where
resonant light--matter coupling gives rise to hybrid states known
as polaritons, possessing both photonic and excitonic character. Following
the pioneering experiments of the Ebbesen group \cite{12HuScGe},
polaritonic chemistry has rapidly developed into a vibrant interdisciplinary
field at the interface of physics and chemistry, offering new strategies
for controlling molecular and material properties. A broad range of
experimental \cite{12HuScGe,15ToBa,16ChNiBe,16Ebbesen,16ThGeSh,16VeGeCh,16ZhChWa,18GrTo,19DaWeKr,19OjChDe,19RoShEr,19VeThNa,24JaReSi}
and theoretical \cite{15GaGaFe,16GaGaFe,16KoBeMu,17FlApRu,17HeSp,18FeGaGa,18PiScCh,18RuTaFl,18SzHaCs_2,18Vendrell,19ReSoGe,19TrSa,21TrSa,22RiHaRo,22MaAkBe,23ScKo_2,23ScKo,23ScSiRu,23Szidarovszky,23Szidarovszky_2,24FiRi,24SoGr,24BaReHo,25Szidarovszky,25FaHaHo,24SaFeGa,18DuMaRi,20LiNiSu,20LiSuNi,19SeNi,21LiNiSu,20MaKrHu,21WePuSc,21Wang,19CaRiZh,19GaClGa,22LiCuSu,22FrGaFe,20FrCoPe,20FrGrPe}
studies have demonstrated that strong coupling can either enhance \cite{23FrCo,24SaFeGa}
or suppress \cite{16GaGaFe} photochemical reaction rates, thereby
affecting processes such as photoisomerization, \cite{18FrGrCo,20FrGrPe}
photodissociation, \cite{20DaKo_2,21ToFe,25SzFaHa} photoionization
\cite{24FaHaCe}, and photoassociation \cite{24CeFe}. Furthermore,
cavity-mediated light--matter interactions can modify charge \cite{19SeNi,20MaKrHu,21WePuSc}
and long-range energy transfer \cite{18DuMaRi,21LiNiSu}, alter absorption
spectra \cite{18SzHaCs_2,21FaHaCe,21FaHaCe_2}, and induce pronounced
nonadiabatic effects \cite{18SzHaCs_2,17CsHaCe_2,19CsKoHa,20GuMu,20GuMu_2,20SzHaVi,21FaMaHu,21SzBaHa,22BaUmFa,22CsVeHa,22FaHaCe,22FiSa,23HaXiHu,24FaCsHa_2}
through ultrafast nonradiative decay pathways.

The strong-coupling regime is reached when the rate of energy exchange
between cavity photons and molecules exceeds the rates of photon loss
and system dephasing. This regime is more readily achieved in plasmonic
nanocavities, whose highly confined electromagnetic modes correspond
to much smaller mode volumes than those of conventional optical cavities.
However, these modes typically exhibit much shorter lifetimes due
to significant photon leakage. Plasmonic modes are hybrid light--matter
excitations associated with collective electronic oscillations in
metals and can be viewed as resonances embedded in the electromagnetic
continuum. Their quantization leads naturally to bosonic polaritonic
modes describing mixed photonic and material excitations. In the long-wavelength
(dipole) approximation, molecule--field interactions in such systems
are dominated by long-range Coulomb interactions. Since these interactions
are unaffected by the Power--Zienau--Woolley transformation \cite{14VuGrDo,20Woolley},
the coupling can be described by the standard $\mathrm{E}\cdot\mathrm{d}$
interaction term without explicitly including the dipole self-energy
contribution.

Coupling between nuclear and electronic motion in polyatomic molecules
gives rise to nonadiabatic phenomena such as conical intersections
(CIs), where degeneracies between electronic potential-energy surfaces
(PESs) strongly influence molecular dynamics, spectroscopy, and topology.
Similar nonadiabatic effects can also be generated by external classical
\cite{08MoSiCe,11HaViSi,13HaViMo,15HaViCe} or quantized electromagnetic
fields \cite{17CsHaCe_2,22FiSa,24FaCsHa_2}. In the presence of intense
laser radiation or the quantized field of optical or plasmonic nanocavities,
light-induced conical intersections (LICIs) may emerge. Although LICIs
exhibit many features analogous to natural CIs, both their location
and coupling strength can be externally controlled through the light
field, providing a powerful route for manipulating molecular properties
and dynamics. 

In the present work, we intend to investigate the light-induced nonadiabatic
dynamics of the $\mathrm{Na_{2}}$ molecule in a lossy cavity. Although
$\mathrm{Na_{2}}$ is a diatomic molecule possessing only a single
vibrational degree of freedom, the inclusion of rotational motion
together with vibration, enables a proper description of light-induced
nonadiabatic effects. In this context, we perform two different types
of numerical simulations:  (i) a treatment where the rotational degree
of freedom is included parametrically by averaging over many fixed
rotational angles, allowing the description of a light-induced avoided
crossing (LIAC) within a one-dimensional (1D) framework; and (ii)
a fully dynamical treatment of the rotational degree of freedom, which
makes it possible to construct the proper two-dimensional (2D) branching
space (BS) and thereby describe the emergence of a LICI, appearing
as a characteristic feature above the BS.

To account for finite photon lifetimes and inherent decoherence, the
quantum dynamics of the cavity--molecule system usually described
by using the density operator ($\varrho$) of the system governed
by the Lindblad master equation (ME). In this formalism, photon losses
and coupling to a dissipative Markovian environment are incorporated,
and the temporal evolution of the system is obtained by propagating
$\varrho$ in time according to the appropriate Lindblad ME \cite{20Manzano,16GaGaFe,20DaKo_2,22FiWeBo,22FaHaVi,24FaCsHa}.
However, cavity leakage can be incorporated in the non-Hermitian time-dependent
Schr\"odinger equation (TDSE) \cite{20FeFrSc,20UlVe,24FaCsHa_2} or
the so-called Stochastic Schr\"odinger equations (SE) \cite{22MaGaBi,22MaGaBi_2,22TrHe}
as well. In the former case the Hamiltonian of the system is augmented
with an imaginary term. The effective non-Hermitian Hamiltonian obtained
in this manner implicitly includes the dissipative effect, which is
then accounted for by a loss of norm of the nuclear wave packet during
time propagation using the TDSE. While the stochastic SE method based
on the Monte Carlo wave packet formalism.

The goals of this paper are two-fold. First, efforts are made to compare
the performance of the three different approaches: the non-Hermitian
TDSE, the stochastic SE, and the Lindblad ME, and analyze the extent
to which these methods yield similar or different results. Second,
we investigate the quantum light-induced nonadiabatic dynamics of
the $\mathrm{Na_{2}}$ molecule in a lossy nano-cavity. Such a study
requires a full 2D treatment, where LICIs can emerge. This enables
us to reveal the differences between the effects of light-induced
avoided crossings (LIACs) obtained within a 1D description, and the
LICIs are appearing in the full 2D framework. 

We note that the corresponding nonadiabatic dynamics induced by classical
laser fields has previously been investigated in this system \cite{11HaViSi,12HaViMo,12HaSiMo}. 

The structure of the paper is organized as follows. Section II. gives
the form of the working Hamiltonian and provides descriptions of the
different dynamical methods (i) Lindblad ME, 
%\textcolor{violet}{ %% Krisztian
(ii) Stochastic SE, (iii)  non-Hermitian TDSE approach
%} 
% Switched the order of non-Hemritian TDSE and stochstic SE, to align with the actual ordering %% Krisztian
and (iv) a brief summary about
the numerical details of the different calculations. In section III.,
results for the different models are presented and discussed, and
conclusions are given in section IV. 

\section*{II. Hamiltonian, Methods, and Computation Protocol}

In the following, a brief overview of the nuclear Hamiltonian governing
the system dynamics is presented. Subsequently, the theoretical approaches
employed to compute the time evolution of the nuclear wave packet,
together with the expressions for the calculated dynamical observables,
are outlined. Numerical details concerning the electronic structure
of the $\mathrm{Na_{2}}$ molecule and the dynamical simulations are
also briefly discussed.

\subsection*{A. The working Hamiltonian}

\begin{figure}
\includegraphics[width=0.4\textwidth]{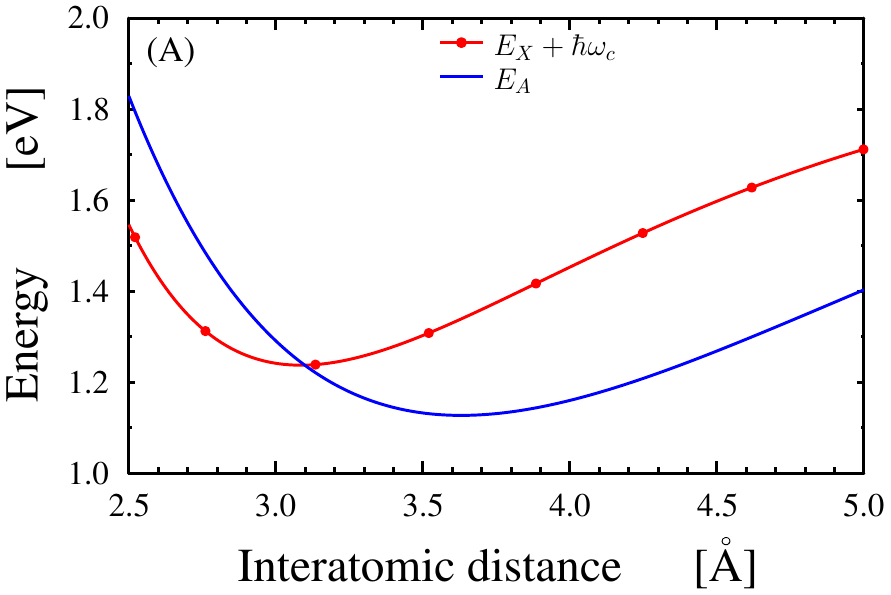}

\includegraphics[width=0.4\textwidth]{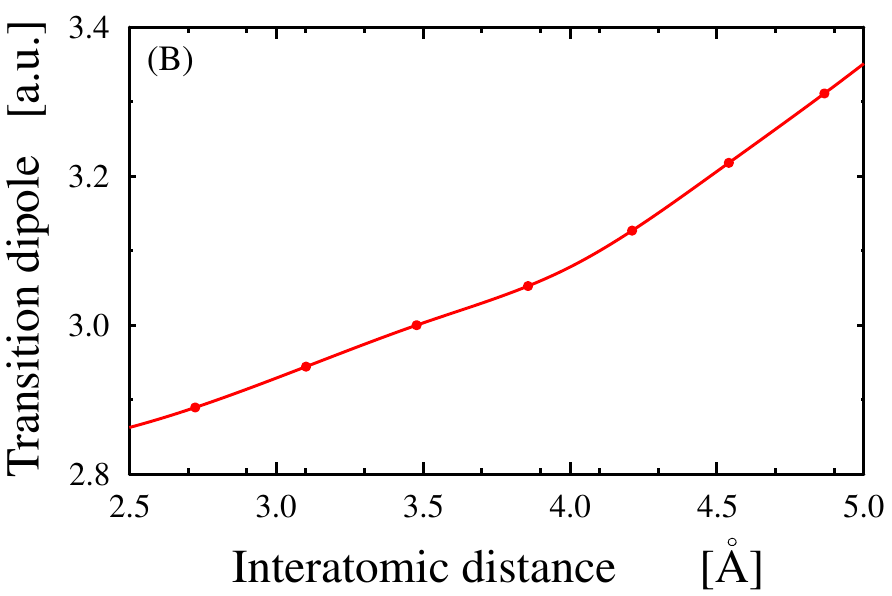}

\includegraphics[width=0.4\textwidth]{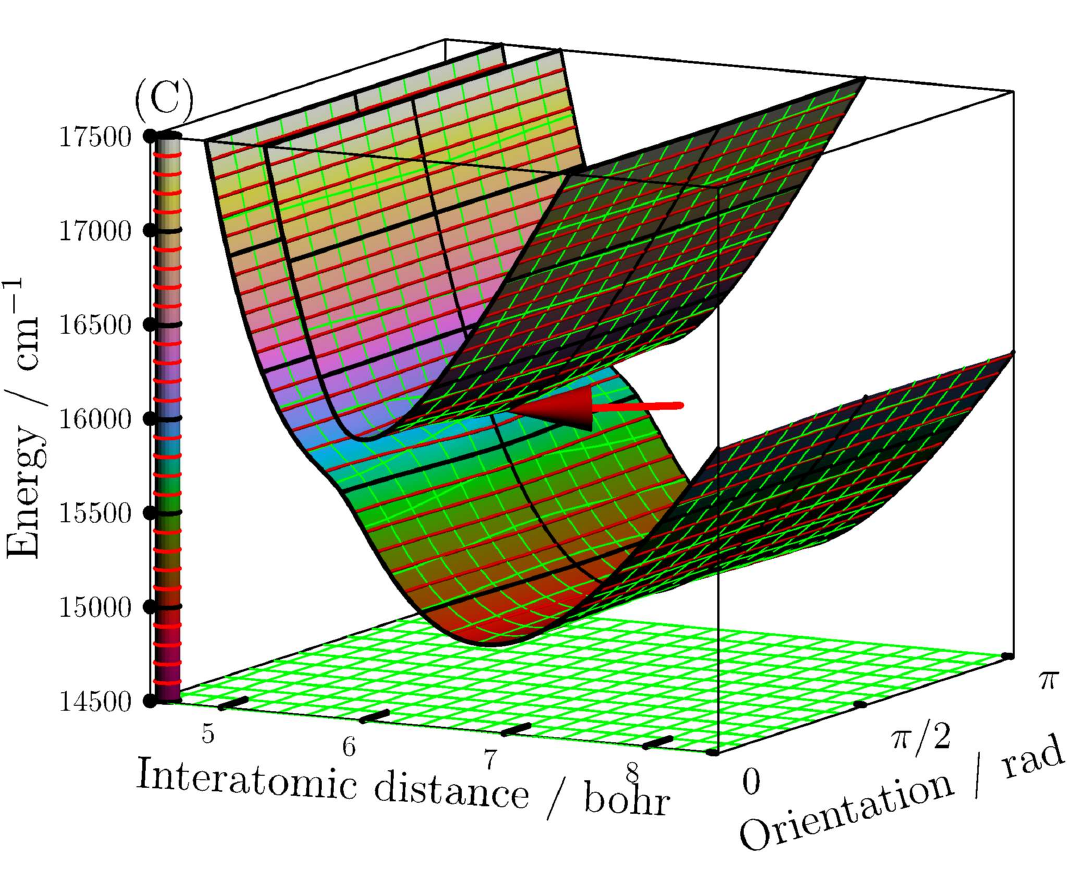}\caption{\label{fig.1} (A) Diabatic potential energy curves of the first excited manifold of the Na$_2$ molecule in a cavity. 
(B) Transition dipole moment (red line) function
between the $\mathrm{X}^{1}\Sigma_{g}^{+}$ and $\mathrm{A}^{1}\Sigma_{u}^{+}$ electronic states. (C) Dressed states potential energy surfaces of the Na$_2$ molecule representing the quantum light-induced conical intersection (LICI). The red arrow marks the location of the LICI.
}

\end{figure}

Let us consider the $\mathrm{Na_{2}}$ molecule as our sample system.
In the calculations we will consider the ground $\mathrm{X}^{1}\Sigma_{g}^{+}$
and the first singlet excited $\mathrm{A}^{1}\Sigma_{u}^{+}$ states
of the molecule, which are coupled by a cavity photon of frequency
$\omega_{c}$. These are characterized by the $\mathrm{V}_{\mathrm{X}}(R)$
and $\mathrm{V}_{\mathrm{A}}(R)$ PESs (see Fig. \ref{fig.1}A).
The corresponding Born Oppenheimer potentials were taken from ref
\cite{93MaMiDu}. 

A molecule interacting with a single cavity mode can be described
by the Hamiltonian, 
\begin{align}
\hat{H}_{cm}=\hat{H}_{\textrm{M}}+\hbar\omega_{\textrm{c}}\hat{a}^{\dag}\hat{a}-g\hat{\vec{d}}\vec{e}(\hat{a}^{\dag}+\hat{a}),\label{eq:operator_H}
\end{align}
 where $\hat{H}_{\textrm{M}}$ is the Hamiltonian of the isolated
molecule, $\hat{a}^{\dag}$ and $\hat{a}$ are the creation and annihilation
operators associated with the cavity mode. $g=\sqrt{\frac{\hbar\omega_{\textrm{c}}}{2\epsilon_{0}V}}$
is the coupling strength parameter, where $\epsilon_{0}$ and $V$
are the permittivity and quantization volume of the cavity, respectively.
$\hat{\vec{d}}$ stands for the transition dipole moment (TDM) operator
\cite{81ZeVeVu} of the molecule and $\vec{e}$ refers to the polarization
vector. In this work, the quadratic dipole self-energy term \cite{25FaHaHo}
is not included in Eq. \ref{eq:operator_H} as it is expected to add
small shifts in the two-state model applied in this work. 

Then the Hamiltonian matrix $\hat{H}_{\textrm{cm}}$ of Eq. \ref{eq:operator_H}
by using rotating wave approximation (RWA) takes the form 
\begin{equation}
\resizebox{0.5\textwidth}{!}{\ensuremath{\hat{H}_{\textrm{cm}}=\begin{bmatrix}\hat{T}+V_{\textrm{X}} & 0 & 0\\
0 & \hat{T}+V_{\textrm{A}} & W_{\textrm{XA}}\\
0 & W_{\textrm{XA}} & \hat{T}+V_{\textrm{X}}+\hbar\omega_{\textrm{c}}
\end{bmatrix}}}\label{eq:matrix_H}
\end{equation}
in the direct product basis $|\alpha,n\rangle=|\alpha\rangle\otimes|n\rangle$
of the electronic states $|\alpha\rangle$ ($\alpha=\textrm{X},\textrm{A}$)
and Fock states $|n\rangle$ ($\alpha=\textrm{0},\textrm{1}$) of
the cavity mode. 
% In Eq. \ref{eq:matrix_H} Felt cumbersome
$\hat{T}=-\frac{\hbar^{2}}{2\mu}\frac{\partial^{2}}{\partial R^{2}}+\frac{\hat{L}^{2}}{2\mu R^{2}}$
is the nuclear (rotational-vibrational) kinetic energy operator, where
$\mu$ is the reduced mass of the two $\mathrm{Na}$ atoms and $\hat{L}^{2}$
is the squared angular momentum operator associated with a diatomic
rotations; $\mathrm{R}$ is the molecular vibrational coordinate.
The cavity-molecule coupling is described by the operator $W_{\textrm{XA}}=-gd_{\textrm{XA}}$. 

The Hamiltonian $\hat{H}_{\textrm{cm}}$ of Eq. \ref{eq:matrix_H}
corresponds to the so-called diabatic representation. Polaritonic
(adiabatic) PESs can be obtained as eigenvalues of the potential energy
part of $\hat{H}_{\textrm{cm}}$ at each nuclear configuration. Of
particular importance for this study is the so-called singly excited
subspace (molecule in its ground electronic state with one photon
and molecule in the excited electronic state with zero photon). This
singly excited subspace accommodates the lower (1LP) and upper (1UP)
polaritonic states, which can be approximately described as superposition
of the states $|\textrm{X},1\rangle$ and $|\textrm{B},0\rangle$,
thus carrying both photonic and excitonic characters. Due to molecular
rotation, the upper and lower adiabatic surfaces are no longer completely
separated; instead, a light-induced conical intersection (LICI) emerges
between them (see Fig. \ref{fig.1}B-\ref{fig.1}C), where the nonadiabatic
couplings become infinitely strong.

In some numerical simulations, $\hat{H}_{\textrm{cm}}$ is supplemented
with a term that describes the interaction of the cavity mode with
a pump laser pulse, that is, 

\begin{equation}
\hat{H}=\hat{H}_{\textrm{cm}}-d_{\textrm{c}}E(t)(\hat{a}^{\dag}+\hat{a})\label{eq:fullH}
\end{equation}
where $d_{\textrm{c}}=1.0~\textrm{au}$ (effective dipole moment of
the cavity mode) and the pump pulse is specified by $E(t)=E_{0}\sin^{2}(\pi t/T)\cos(\omega_{laser} t)$
for $0\le t\le t_{max}$ and $E(t)\equiv0$ otherwise. $E_{0}$, $T$
and $\omega_{laser}$ denote the amplitude, length and carrier angular
frequency of the pump pulse, respectively. $E_{0}$ can be converted
into laser intensity by the formula $I=c\epsilon_{0}E_{0}^{2}/2$
($c$ is the speed of light in vacuum and $\epsilon_{0}$ is the vacuum
permittivity). 

The Hamiltonian in Eq. \ref{eq:operator_H} assumes an infinite lifetime
of the cavity field excitation. In practice, however, finite photon
lifetimes cannot be neglected, particularly in plasmonic nanocavities,
which are inherently lossy, yet enable access to the single-molecule
strong coupling regime. Since the cavity lifetime is often much shorter
than the molecular dynamical timescales, dissipative effects must
be explicitly taken into account. To incorporate finite photon lifetimes
and associated decoherence, we describe and compare the nonadiabatic
light-induced quantum dynamics of the $\mathrm{Na_{2}}$ molecule
applying three different methods. These are: the density operator
$\hat{\rho}$ based approach, in which the time evolution is governed
by the Lindblad ME, the non-Hermitian TDSE as well as the so-called
Stochastic SE approaches. The results obtained using the three approaches
provide an opportunity to compare the performance of the different
methods through the investigation of light-induced dissipative nonadiabatic
quantum dynamics.

\subsection*{B. Applied methods}

\subsubsection*{Lindblad master equation}

Usually, the dynamics of a system coupled to a Markovian environment
described by the Lindblad ME, \cite{20Manzano} which propagates the
density operator $\hat{\rho}$ pertaining to the system
%
%\begin{equation}
%    \label{eq:Lindblad}
%    \frac{\partial\hat{\rho}}{\partial t}=-\frac{\textrm{i}}{\hbar}
%    [\hat{H},\hat{\rho}] + \gamma_\textrm{c} \left(\hat{a} \hat{\rho} \hat{a}^\dag - \frac{1}{2} \left\{ \hat{a}^\dag \hat{a} \hat{\rho} %\right\}\right)
%\end{equation}
%
%
 \begin{equation}
 \frac{\partial\hat{\rho}}{\partial t}=-\frac{\textrm{i}}{\hbar}[\hat{H},\hat{\rho}]+\gamma_{\textrm{c}}\hat{a}\hat{\rho}\hat{a}^{\dag}-\frac{\gamma_{\textrm{c}}}{2}(\hat{\rho}\hat{N}+\hat{N}\hat{\rho})\label{eq:Lindblad}
 \end{equation}
%
%% \textcolor{violet}{ %% Krisztian
%% I think we should avoid the $\hat{N}=\hat{a}^{\dag}\hat{a}$ notation in the ME, because it's not the standard form. Moreover, $\hat{a}^{\dag}\hat{a}$ does not appear here as a number operator, rather a product of a Lindbad discipator. Eq 4. is true, but Conceptually not general. The general form, applied to our case of single dissipation channel in the cavity would look like the one above.}
%
where $\hat{\rho}$ denotes the density operator and $\hat{N} = \hat{a}^{\dag}\hat{a}$
% where $\hat{\rho}$ denotes the density operator and $\hat{N}=\hat{a}^{\dag}\hat{a}$
%
%% We can flip the notation below, as $\hat{a}^{\dag}\hat{a} = \hat{N}$, that way we can still emphasize that it is, in fact the photon number OP.
% 
% the DM was already identified at the top
is the photon number operator associated with the cavity mode. Moreover,
$\gamma_{\textrm{c}}$ denotes the cavity decay rate which is equivalent
to a lifetime of $\tau=1/\gamma_{\textrm{c}}$. 

The Lindblad ME can be regarded as a deterministic approach, since
it does not rely on stochastic processes and, for a given system,
always produces the same result.
%
%% I modified this part to match the left out part.
%
This contrasts with the stochastic SE \cite{22MaGaBi,22MaGaBi_2,22TrHe}
%
%% I would move the text below with green to the stochastic SE part.
%
, where a single realization (quantum trajectory) consists of deterministic evolution interrupted by random quantum jumps. The density matrix in Eq. \ref{eq:Lindblad} can be interpreted as the ensemble average over infinitely many quantum trajectories, $\rho=\mathrm{E}(|\psi\rangle\langle\psi|)$, where $\mathrm{E}$ denotes the classical expectation value.
Solving Eq. \ref{eq:Lindblad} represents a considerable numerical challenge because
the propagation of the density matrix scales as $\mathcal{O}(N^{4})$
with system size. Consequently, even our relatively simple $\mathrm{Na_{2}}$
model becomes computationally demanding due to the spatially resolved
molecular degrees of freedom.

\subsubsection*{Stochastic Schr\"odinger equation}

Continuous monitoring of the environment gives rise to a stochastic
process $\psi(t)$ in 
%the %% Krisztian
Hilbert space, where the random variable is
the state vector of the reduced system itself \cite{22MaGaBi,22MaGaBi_2,22TrHe}.
This monitoring acts as an indirect and selective measurement, leading
to stochastic quantum jumps in the system wave function. In the diffusion
limit, this framework yields the working form of the stochastic SE

\begin{align}
d\psi(t)=-iH\psi(t)-\sum_{n}\left(\frac{S_{n}^{\dag}S_{n}}{2}-\frac{e_{n}}{2}S_{n}+\frac{e_{n}^{2}}{8}\right)+\sum_{n}\left(S_{n}-\frac{e_{n}}{2}\right)\psi(t)dW_{n},\label{Eq:SSE}
\end{align}
where $S_{n}$ are the stochastic collapse operators, $dW_{n}$ is
the Wiener increment, and $e_{n}=\langle\psi(t)|S_{n}+S_{n}^{\dag}|\psi(t)\rangle$.
A single solution of Eq. \ref{Eq:SSE} is referred to as a quantum
trajectory, representing one realization of the stochastic process
$\psi(t)$. 

The Lindblad ME and the stochastic SE provide two different descriptions
of the same physical system. The corresponding stochastic equation
is commonly referred to as an unraveling of the master equation, since
the density matrix can be reconstructed from the ensemble average
of the stochastic state vectors, $\rho=E(|\psi\rangle\langle\psi|)$.
The stochastic wave function dynamics describes the evolution of the
system under indirect, selective measurements performed on the environment.
Averaging over many realizations of the stochastic process yields
the non-selective dynamics --- equivalent to tracing over the environmental
degrees of freedom --- and thus recovers the corresponding equation
of motion for $\hat{\rho}$.
%\textcolor{violet}{ %% Krisztian
%The stochastic SE propagates the stochastic state vector over many realizations, therefore the computational effort scales as $\mathcal{O}(M \cdot N^2)$ with $M$ being the number of the considered quantum trajectories. For $M < N^2$ it is more efficient than Eq. \ref{eq:Lindblad}, even before the utilization of the more straightforward parallelization.
%}
%% The Lindblad part mentions the computational complexity, I think the same should be mentioned here. %% Krisztian

\subsubsection*{Non-Hermitian Schr\"odinger equation}

Dissipation during the time evolution of an open quantum system can be described by the non-Hermitian TDSE, \cite{20UlVe,20FeFrSc,24FaCsHa}
which takes the form

\begin{equation}
\textrm{i}\hbar\frac{\partial|\psi\rangle}{\partial t}=\Bigl(\hat{H}-\textrm{i}\frac{\gamma_{\textrm{c}}}{2}\hat{N}\Bigr)|\psi\rangle.\label{eq:Schrodinger}
\end{equation}
As shown in Ref. \cite{20FeFrSc} the Lindblad ME becomes equivalent
to the TDSE of Eq. \ref{eq:Schrodinger} when the $\gamma_{\textrm{c}}\hat{a}\hat{\rho}\hat{a}^{\dag}$
term is omitted. This term accounts for incoherent transitions $|\alpha(n+1)\rangle\rightarrow|\alpha n\rangle$
($\alpha$ labels molecular electronic states $\mathrm{X}$ and $\mathrm{A}$),
thereby compensating for population losses in states with $n>0$ and
ensuring conservation of the total density-matrix norm. In contrast,
within the non-Hermitian TDSE framework, the wave function norm decreases
over time due to the term $-\textrm{i}\frac{\gamma_{\textrm{c}}}{2}\hat{N}$.
The performance of the non-Hermitian TDSE approach
%\textcolor{violet}{ %% Krisztian
%, if there is only one dissipation channel,
%}
%% In general, it's not true. When there's only 1 dissipation chanel, it OK, according to the cited article. If there's a CAP it's definitely not true. In a sense, a CAP can be regarded as a dissipation channel, as the total energy of the system decreases as the CAP does its thing.
can be improved
by renormalizing the wave function at each propagation step \cite{24FaCsHa},
similarly to the procedure employed in the stochastic SE method. Furthermore,
the TDSE propagates a state vector $|\psi\rangle$ rather than a density
matrix $\hat{\rho}$
%\textcolor{violet}{ %% Krisztian
%or multiple quantum trajectories
%}
%% It's also more efficient than the SSE
, which substantially reduces the computational
cost, especially in systems involving several nuclear degrees of freedom.
Consequently, identifying the regime in which the Schr\"odinger TDSE
provides a reliable approximation is of considerable practical importance.

\subsection*{C. Numerical details}

In the following, the nuclear dynamics governed by the Hamiltonian
in Eq. \ref{eq:operator_H} is investigated using the three different
methods outlined below. For these calculations, we employ the QuTiP
\cite{12JoNaNo,Johansson_2013,26LaGiMe} and cuQuantum \cite{23BaChCl}
software packages. Specifically, QuTiP is used to solve the non-Hermitian
and stochastic SEs, as well as the Lindblad ME for  1D cases, while
the 2D Lindblad ME is propagated using the cuQuantum framework.

In the numerical calculations, the vibrational coordinate was represented
using a sine discrete-variable representation (DVR). Rotational motion
was expanded in $\mathcal{L}^{2}$-normalized associated Legendre
functions with $m=0$. In all calculations, the initial state was
taken as a separable pure state. The vibrational degree of freedom
was prepared in the vibrational ground state of the ground-state electronic
PES, and the rotational degree of freedom in the $l=0$ state. The
initial electronic and photonic states were varied to probe regimes
in which the three different methods agree (see above).

Throughout this work, we calculate two dynamical quantities: the excited-state
population and the mean photon number. For wave function-based methods,
the excited-state population is calculated as the norm of the wave
packet associated with the excited electronic state,

\begin{equation}
P_{\mathrm{e}}(t)=\langle\varPsi_{e}(t)\mid\varPsi_{e}(t)\rangle\label{eq:Popwave}
\end{equation}
 while the mean photon number is given by

\begin{equation}
\langle N\rangle(t)=\langle\psi(t)\mid\hat{a}^{\dagger}\hat{a}\mid\psi(t)\rangle.\label{eq:numberwave}
\end{equation}
In the stochastic Schr\"odinger equation (SSE) approach, these quantities
are first evaluated for each individual quantum trajectory and subsequently
averaged over the ensemble of trajectories to obtain the corresponding
expectation values. For methods based on the density operator formalism,
the excited-state population is computed as 

\begin{equation}
P_{\mathrm{e}}(t)=\mathrm{Tr}\left[\left|e\right\rangle \left\langle e\right|\hat{\rho}(t)\right]\label{eq:popro}
\end{equation}
 whereas the mean photon number is given by

\begin{equation}
\langle N\rangle(t)=\mathrm{Tr}\left[\hat{a}^{\dagger}\hat{a}\hat{\rho}(t)\right].\label{eq:numro}
\end{equation}
 These observables provide direct information about the molecular
excitation dynamics and the occupation of the cavity mode, respectively.

\section*{III. Results and Discussion}

\begin{figure}
\includegraphics[width=0.5\textwidth]{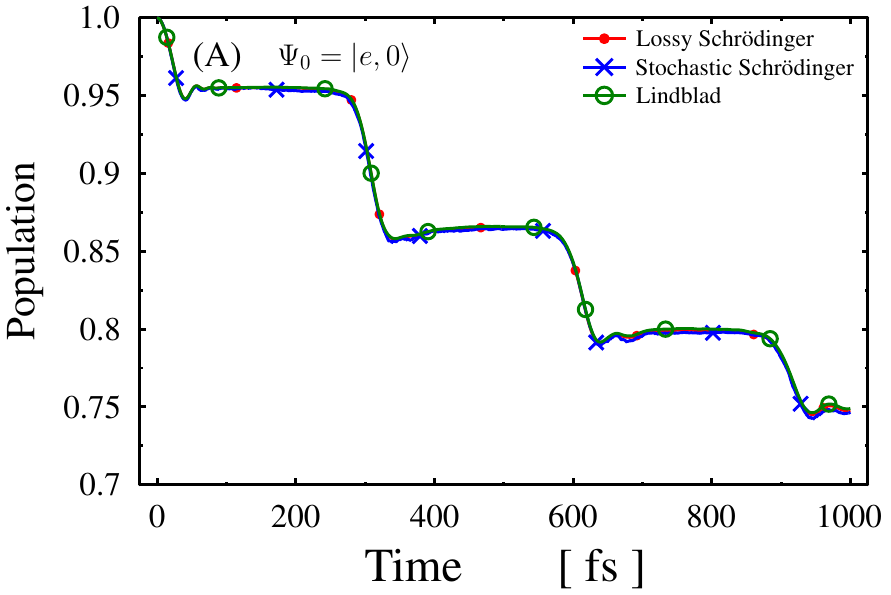}\includegraphics[width=0.5\textwidth]{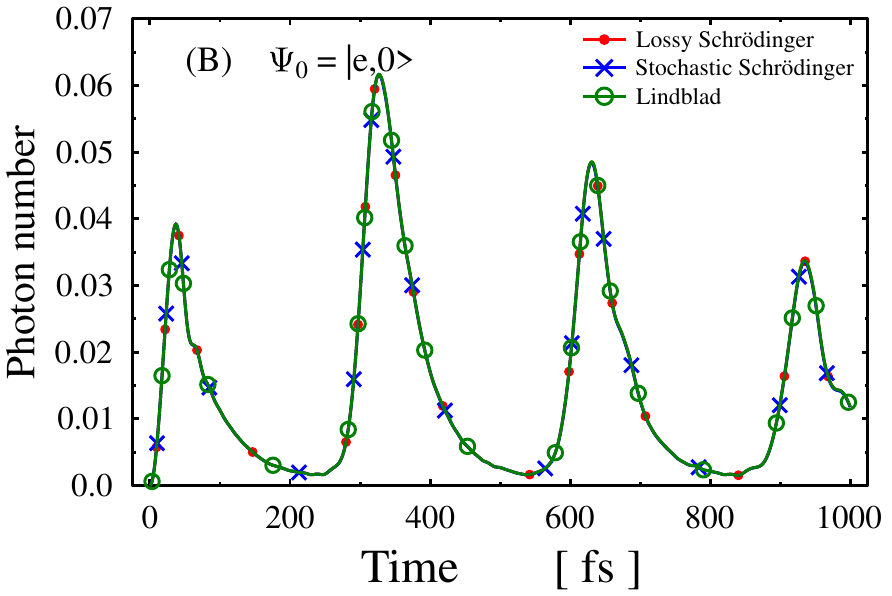}

\includegraphics[width=0.5\textwidth]{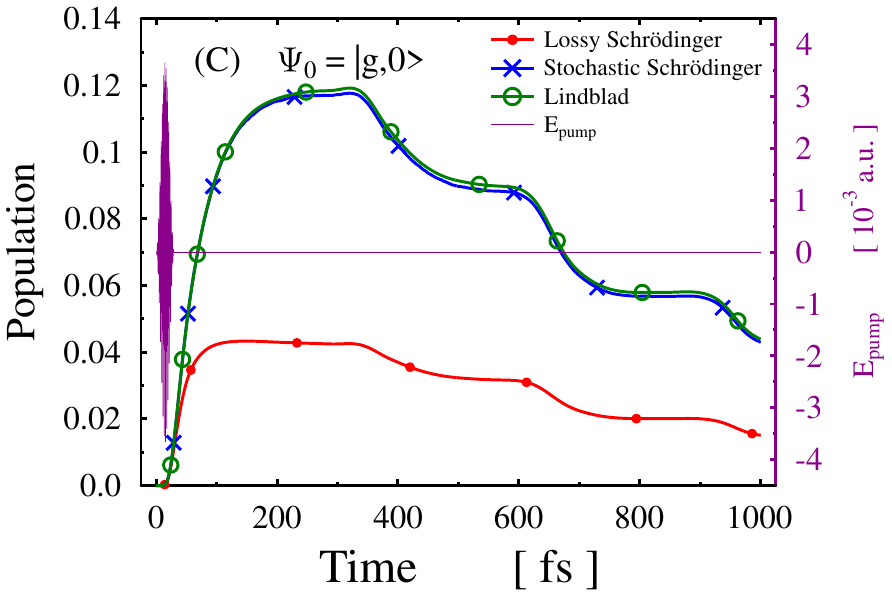}\includegraphics[width=0.5\textwidth]{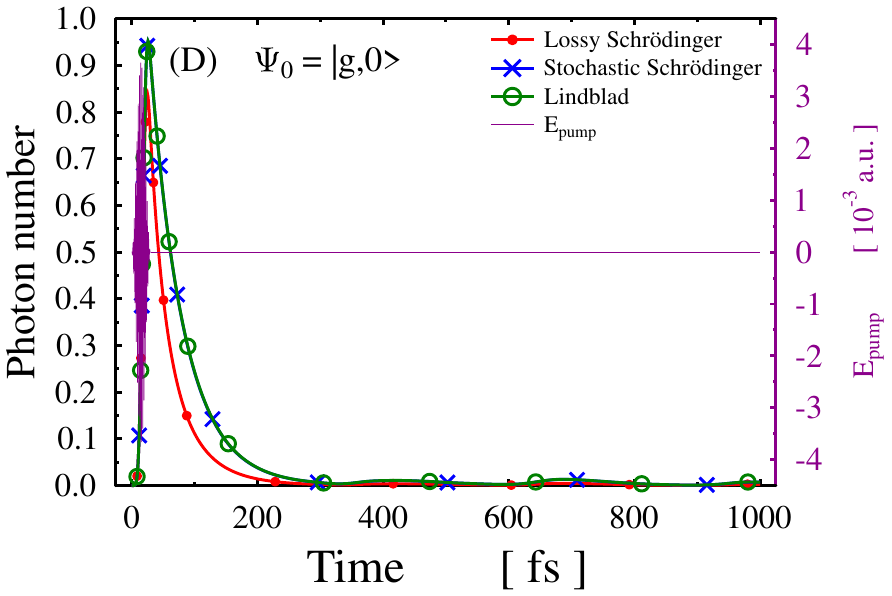}

\caption{\label{fig.2} Population dynamics of the excited molecule and mean
photon number in the 1D framework in a lossy cavity ($g=8\times10^{-5}a.u.,$
$\gamma_{c}=4\times10^{-4}a.u.$ and $\omega_{cav}=1.968eV$). The
dynamics is initiated from the $|e,0\rangle$ state (panels A and
B) and from the $|g,0\rangle$ state pumping the population using laser
pulse as $I=4.74\times10^{11}w/cm^{2},$ $T=30\,\mathrm{fs}$ and
$\omega_{laser}=2eV$ (panels C and D). Three different methods are
used. Lindblad ME results (line with circle), Stochastic SE curve (line
with cross), and non-Hermitian TDSE curve (line with dots). }
\end{figure}

\begin{figure}
\includegraphics[width=0.5\textwidth]{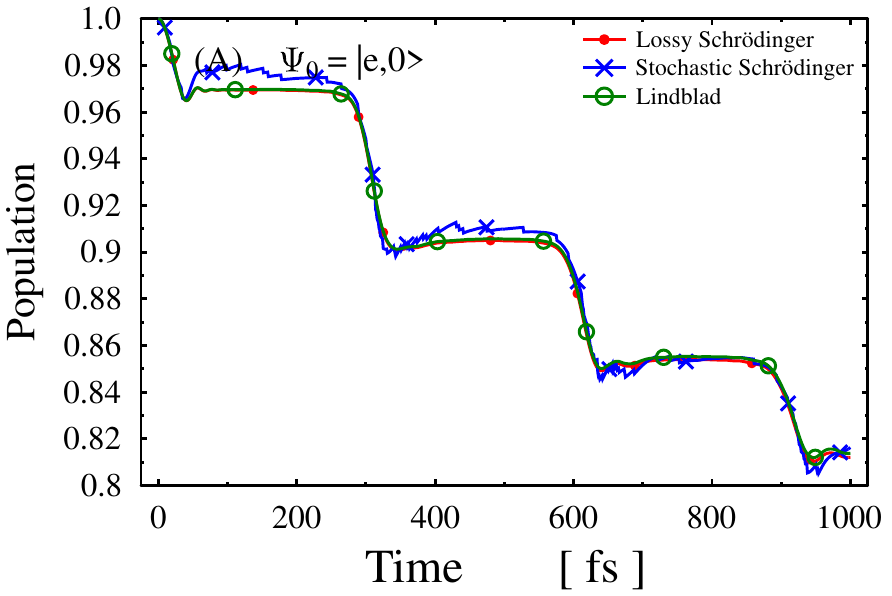}\includegraphics[width=0.5\textwidth]{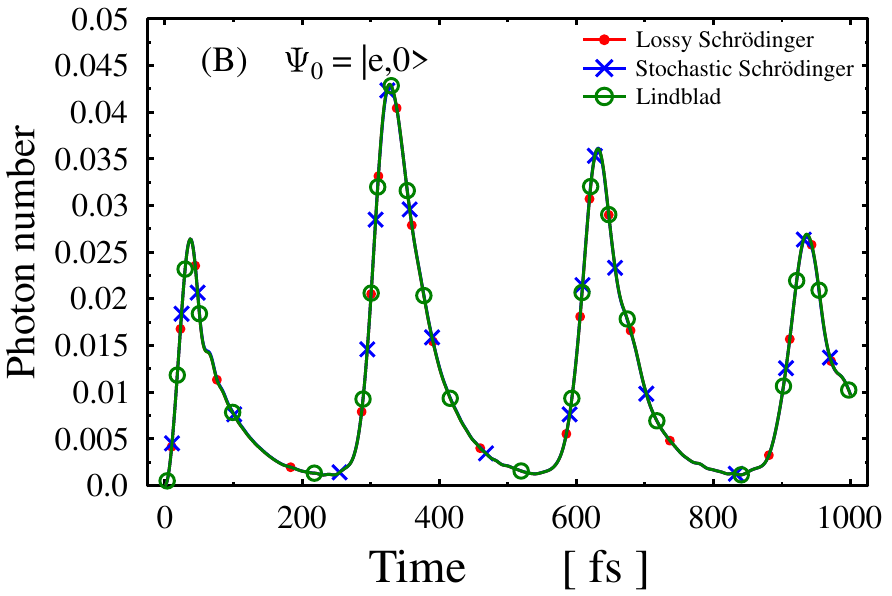}

\includegraphics[width=0.5\textwidth]{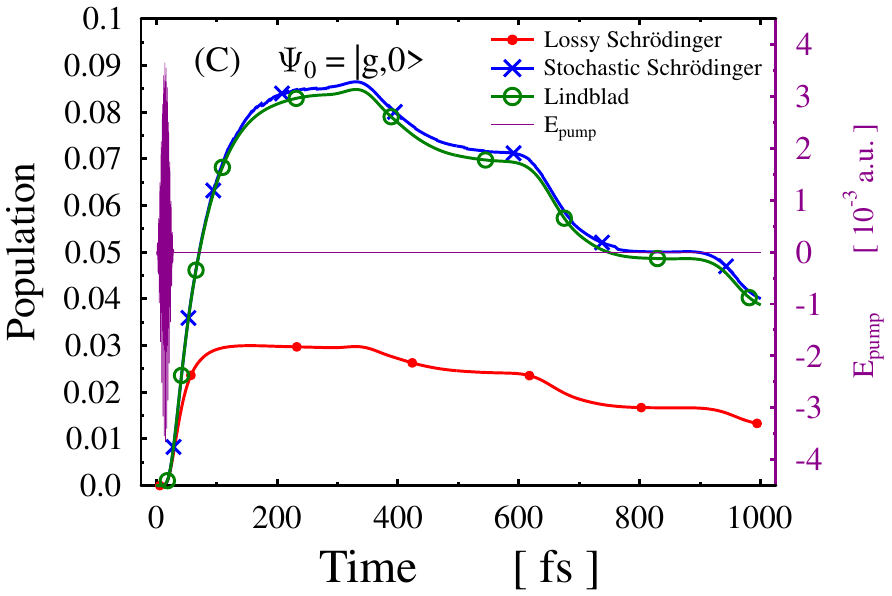}\includegraphics[width=0.5\textwidth]{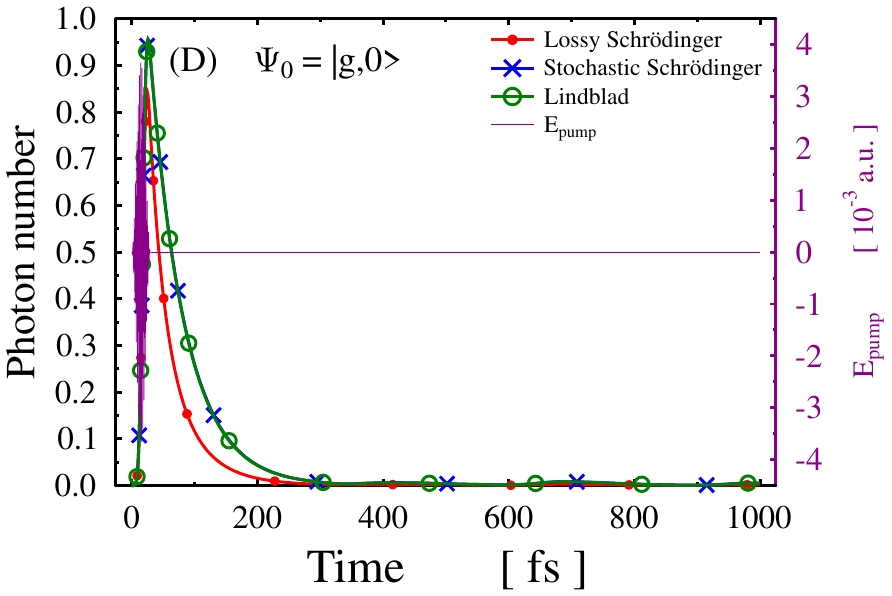}

\caption{\label{fig.3} Population dynamics of the excited molecule and mean
photon number in the 2D framework in a lossy cavity ($g=8\times10^{-5}a.u.,$
$\gamma_{c}=4\times10^{-4}a.u.$ and $\omega_{cav}=1.968eV$). The
dynamics is initiated from the $|e,0\rangle$ state (panels A and
B) and from the $|g,0\rangle$ state pumping the population to the
excited polaritonic manifold using a laser pulse as $I=4.74\times10^{11}w/cm^{2},$
$T=30\,\mathrm{fs}$ and $\omega_{laser}=2eV$ (panels C and D). Three
different methods are used. Lindblad ME results (line with circle),
Stochastic SE curve (line with cross) and non-Hermitian TDSE curve
(line with dots). }
\end{figure}

Having outlined the computational protocol and the methods employed
in this work, we now turn to the presentation of the numerical results.
We focus on the time evolution of the excited state population of
the $\mathrm{Na}_{2}$ molecule and the mean photon number (photon
emission signal) over the time interval $\left\{ t=0-1000\,fs\right\} $.
We will consider different computational protocols: the initial nuclear
wave functions are prepared in the $|e,0\rangle$, \textbar$e,1\rangle$
and \textbar$g,2\rangle$ states of the molecule--cavity system,
as well as a fraction of the ground-state population is promoted to
the excited state of the molecule-cavity system by means of an external
laser pulse applying several different pulse lengths. We investigate
and discuss how the temporal evolution of the considered quantities
depends on both the computational method employed and the choice of
initial conditions for realistic values of the cavity coupling strength
and cavity loss rate. Laser pulses with durations of $T=15,30\,\textrm{and}\,45\,\mathrm{fs}$
are used throughout the study. As for the cavity parameters, the coupling
strengths and cavity loss rate are chosen as ($g=8\cdot10^{-5}a.u.$
; $g=5\times10^{-4}a.u.)$ and $(\gamma_{c}=4\cdot10^{-4}a.u.)$,
respectively. The latter corresponds to a photon lifetime of $\tau_{c}=61\,\mathrm{fs}$.
Such a photon lifetime is considered experimentally realistic for
present-day plasmonic cavities and nanophotonic structures \cite{25CsSzVi}. 

\begin{figure}
\includegraphics[width=0.5\textwidth]{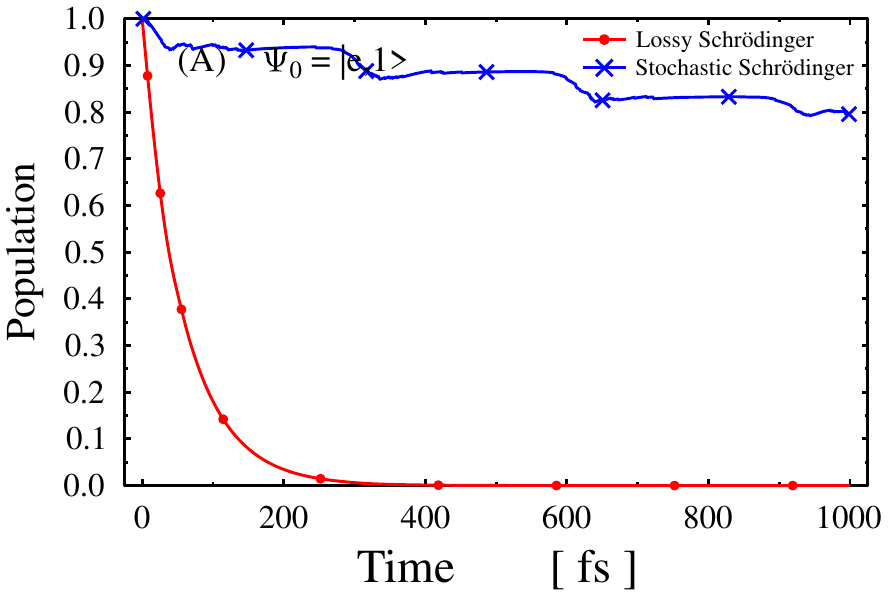}\includegraphics[width=0.5\textwidth]{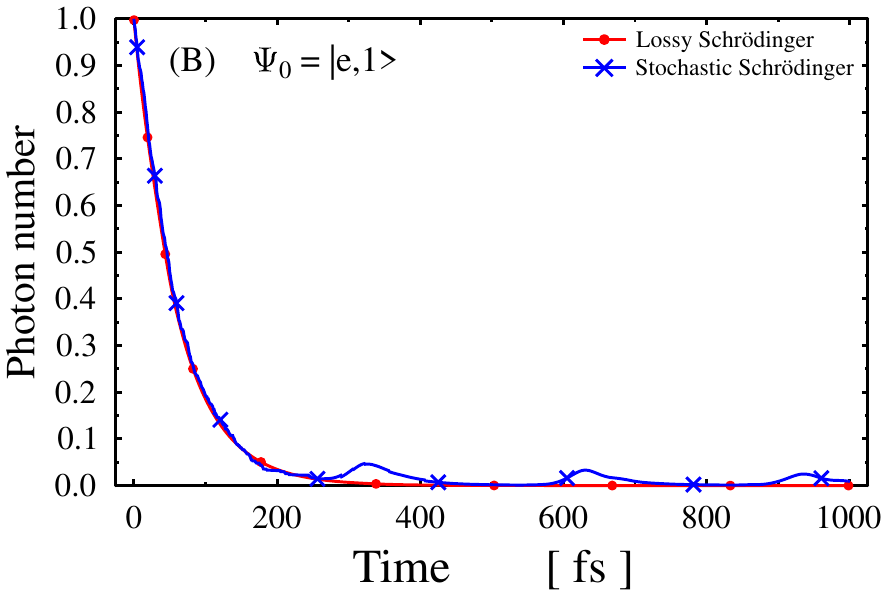}

\includegraphics[width=0.5\textwidth]{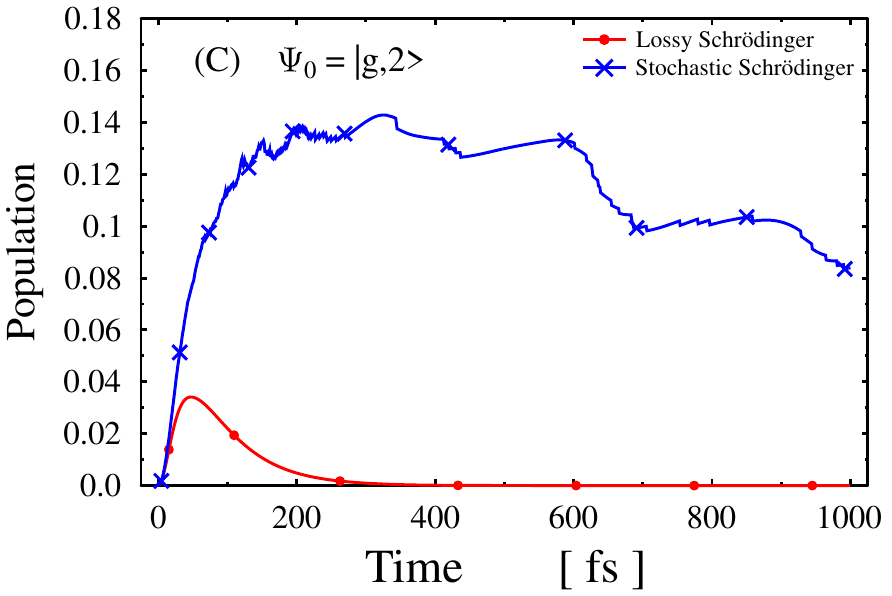}\includegraphics[width=0.5\textwidth]{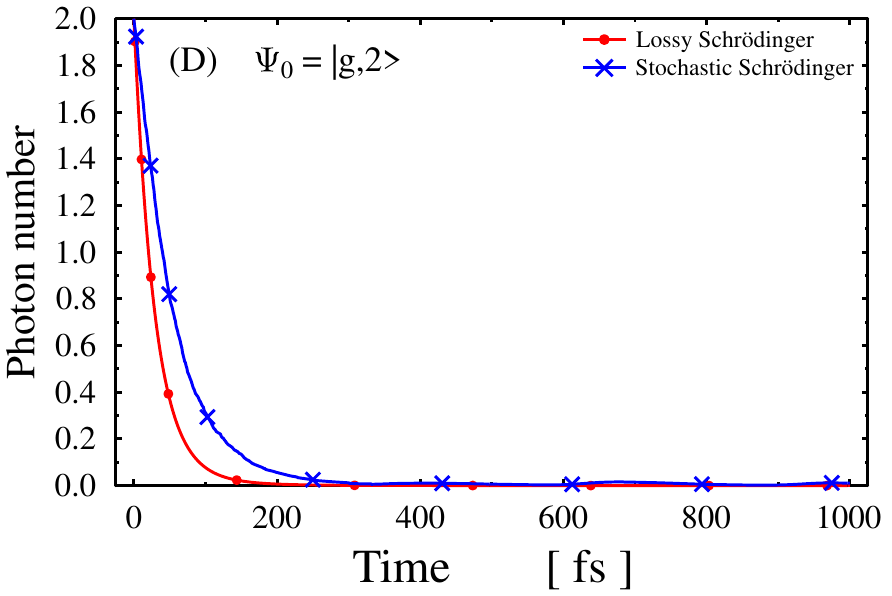}

\caption{\label{fig.4 } Population dynamics of the excited molecule and mean
photon number in the 2D framework in a lossy cavity ($g=8\times10^{-5}a.u.,$
$\gamma_{c}=4\times10^{-4}a.u.$and $\omega_{cav}=1.968eV$). The
dynamics is initiated from the $|e,1\rangle$ state (panels A and
B) and from the $|g,2\rangle$ (panels C and D). Two different methods
are used. Stochastic SE curve (line with cross) and non-Hermitian
TDSE curve (line with dots). }
\end{figure}

\begin{figure}
\includegraphics[width=0.5\textwidth]{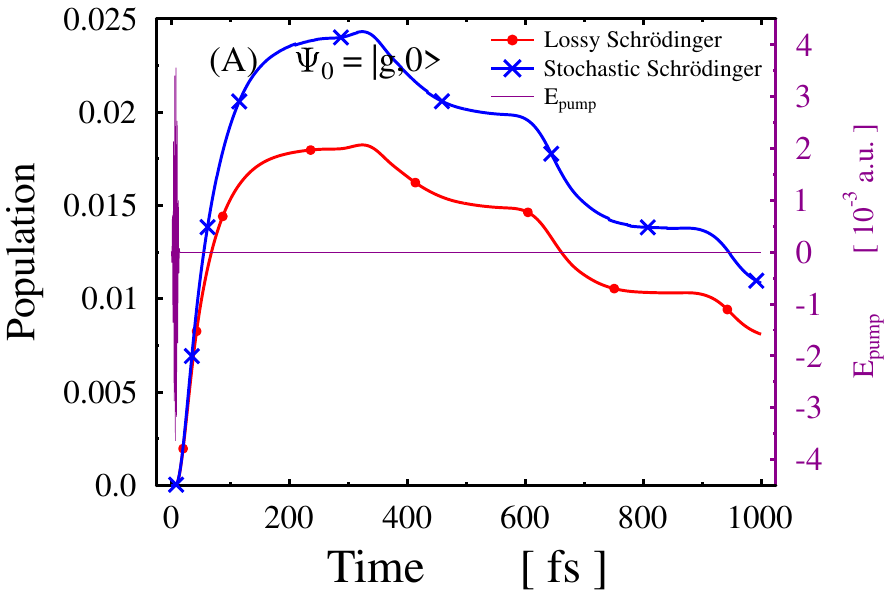}\includegraphics[width=0.5\textwidth]{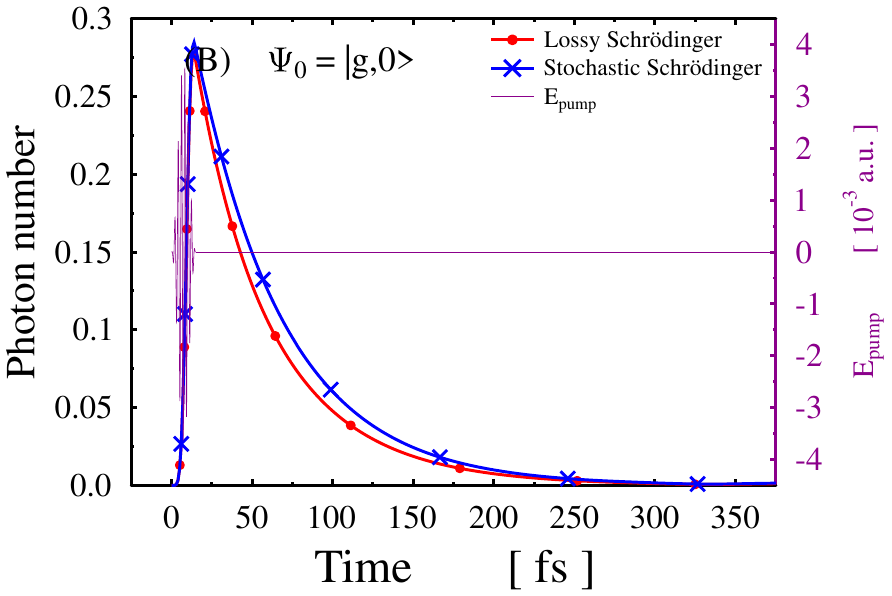}

\includegraphics[width=0.5\textwidth]{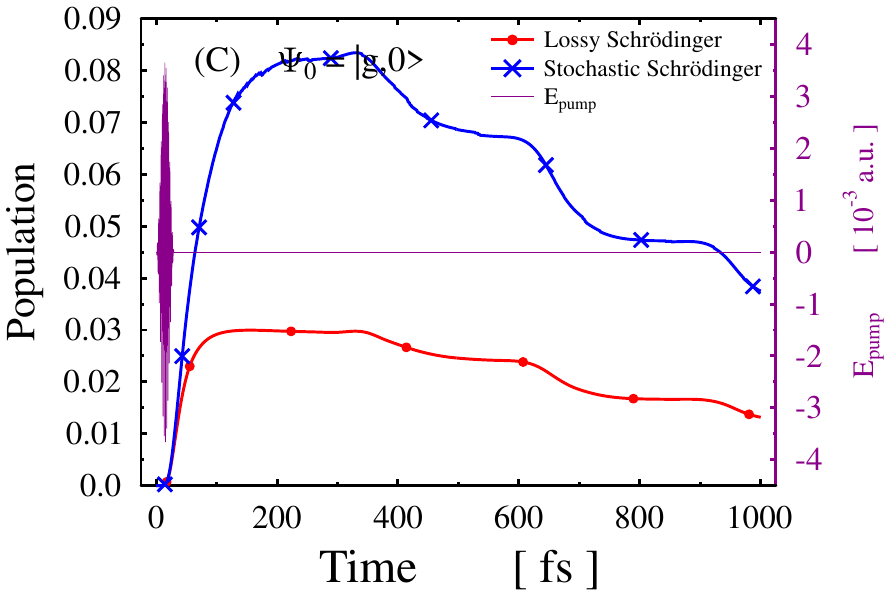}\includegraphics[width=0.5\textwidth]{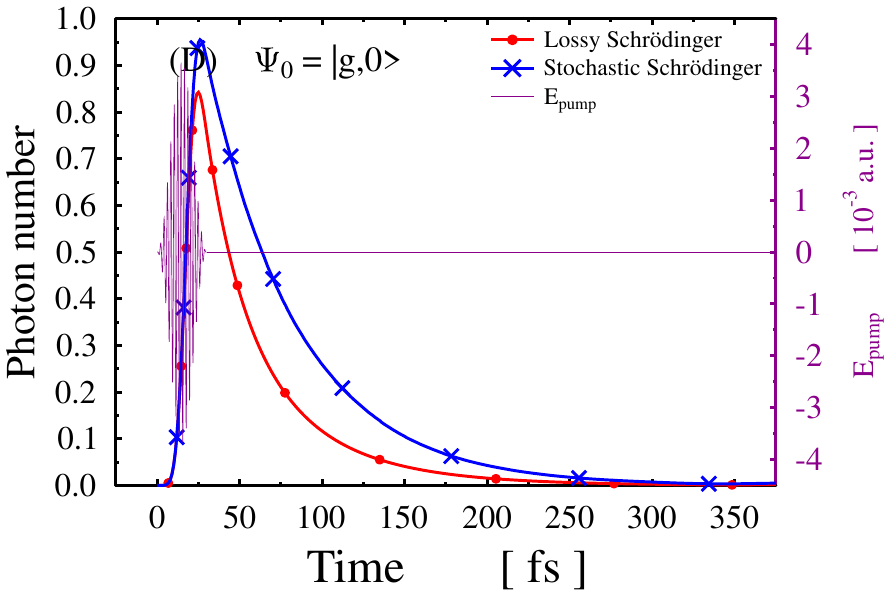}

\includegraphics[width=0.5\textwidth]{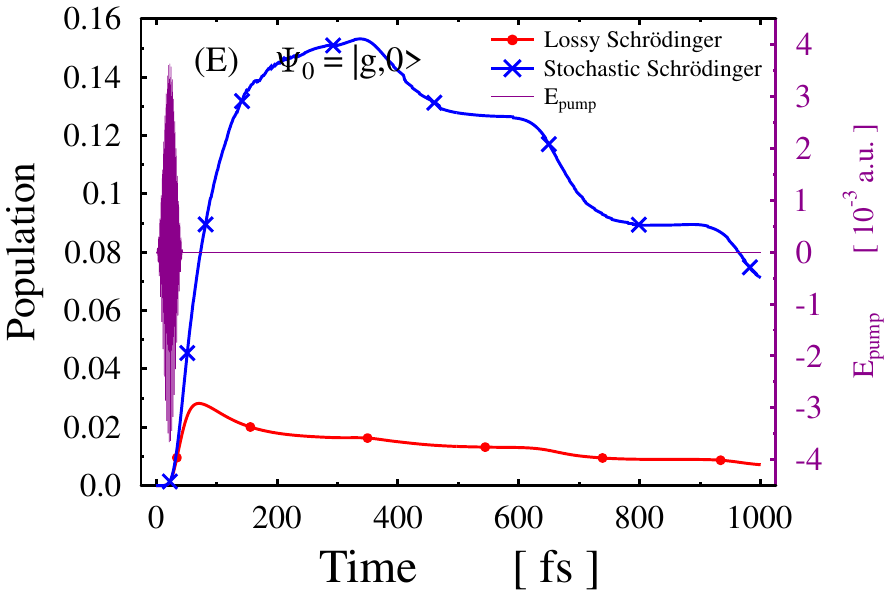}\includegraphics[width=0.5\textwidth]{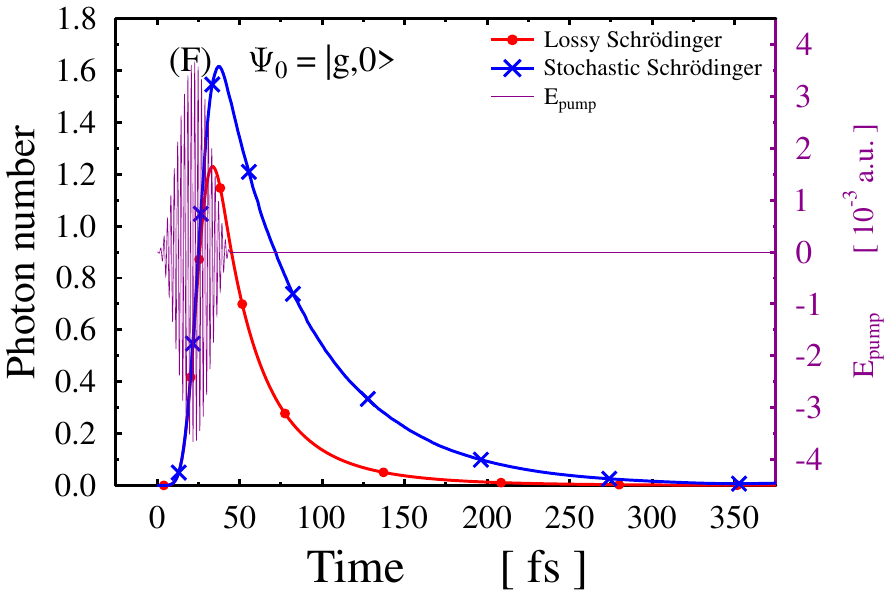}

\caption{\label{fig.5 } Population dynamics of the excited molecule and mean
photon number in the 2D framework in lossy cavity ($g=8\times10^{-5}a.u.,$
$\gamma_{c}=4\times10^{-4}a.u.$ and $\omega_{cav}=1.968eV$). The
dynamics is initiated from the $|g,0\rangle$state pumping the population
to the excited polaritonic manifold using laser pulses as ($I=4.74\times10^{11}w/cm^{2}$
and $\omega_{laser}=2eV$ ) with length $T=15\,\mathrm{fs}$ (panels
A and B), $T=30\,\mathrm{fs}$ (panels C and D) and $T=45\,\mathrm{fs}$
of (panels E and F). Two different methods are used. Stochastic SE
curve (line with cross) and non-Hermitian TDSE curve (line with dots). }
\end{figure}

We begin by comparing the three computational approaches introduced
above. Both 1D and 2D calculations are performed. In the 1D treatment,
only the vibrational degree of freedom of the molecule is considered
as a dynamical variable, while the rotational angle, defined as the
angle between the cavity-photon polarization vector and the molecular
axis, is treated as a fixed parameter. Calculations are carried out
for a range of fixed angles, and the final quantities are obtained
by averaging over these orientations. In contrast, the 2D calculations
explicitly include the rotational degree of freedom as an additional
dynamical variable. These calculations are important not only because
they account for the torque exerted by the electric field on the molecule,
but also because light-induced nonadiabatic effects in diatomic molecules
can only be properly described within such a multidimensional framework
\cite{24FaCsHa_2}. Fig. \ref{fig.2}A and Fig. \ref{fig.2}B present
1D results obtained by initiating the dynamics from the $|e,0\rangle$
state of the molecule-cavity system. Panel (A) shows the excited-state
population, while panel (B) displays the mean photon number. It is
apparent that all three methods yield nearly identical results for
both physical quantities. During the dynamics, the nuclear wave packet
oscillates back and forth on the excited polaritonic state of the
coupled molecule-cavity system, and the population curve closely follows
this motion. Due to the mixed exciton-photon character of the polaritonic
state, certain regions of it are predominantly excitonic. When the
wave packet resides in these regions, the population curve remains
nearly constant. In contrast, when the wave packet enters regions
with strong photonic character, the population shown in Fig. \ref{fig.2}A
decreases rapidly. These events correspond directly to the peaks observed
in the mean photon number shown in Fig. \ref{fig.2}B. It is noteworthy
that the photonic population remains relatively small throughout the
entire dynamics, exceeding neither a few percent nor reaching values
comparable to the molecular excited-state population. In fact, it
is more than an order of magnitude smaller than the molecular population
during most of the propagation. As a consequence, the effective lifetime
of the coupled cavity-molecule system is considerably longer than
the cavity photon lifetime of $\tau_{c}=61\,\mathrm{fs}$ associated
with the chosen cavity loss rate. Since photons can leave the system
only when the photonic component of the polaritonic state is populated,
cavity losses are effective only during relatively short time intervals.
Similar behavior was reported in our previous study of a lossy atom-cavity-molecule
system \cite{25CsSzVi}. We now turn to the results presented in Fig.
\ref{fig.2}C and Fig. \ref{fig.2}D. In these cases, the initial
condition is prepared by exciting the population from the $|g,0\rangle$
state of the cavity-molecule system to the excited polaritonic state
using a $T=30\,\mathrm{fs}$ laser pulse. The subsequent dynamics
is then initiated from the resulting excited-state wave packet. It
is immediately apparent that the results obtained with the non-Hermitian
TDSE differs from those produced by the Lindblad ME and the stochastic
SE. The discrepancy is particularly pronounced in Fig. \ref{fig.2}C..
The origin of this difference is that the non-Hermitian TDSE approach
does not repopulate the ground polaritonic state through the incoherent
decay process. In contrast, both the Lindblad ME and the stochastic
SE explicitly account for this mechanism. When the pulse duration
becomes comparable to or exceeds the cavity lifetime, the incoherent
decay term, ($\gamma_{\textrm{c}}\hat{a}\hat{\rho}\hat{a}^{\dag})$,
begins to repopulate the ground polaritonic state before the laser
pulse has completely vanished. A population that has returned to the
ground polaritonic state can subsequently be re-excited by the laser
field, resulting in a more efficient transfer of population to the
excited polaritonic manifold than in the non-Hermitian TDSE approach.
Another striking feature is the rapid decay of the mean photon number.
It is because the laser pulse directly generates a substantial photonic
component, the cavity losses become much more effective than in the
scenario discussed in Fig. \ref{fig.2}B. Consequently, photon loss
occurs on a significantly faster timescale. The molecular population
displayed in Fig. \ref{fig.2}C exhibits an overall structure similar
to that observed in Fig. \ref{fig.2}A. However, the dynamics starts
from a substantially smaller excited-state population, since the laser
pulse transfers only about 10\% of the initial ground-state population
into the excited polaritonic manifold.

We now proceed to the analysis of the 2D simulations. The results
shown in Fig. \ref{fig.3} correspond to the same computational protocols
discussed in Fig. \ref{fig.2}. However, in the present case, the molecular
rotational degree of freedom is treated as an explicit dynamical variable
rather than as a fixed parameter. Consequently, both vibrational and
rotational degrees of freedom are incorporated into the quantum dynamical
description. Our primary aim is to compare the performance
of the three computational approaches. In this regard, the results
obtained from the 2D simulations exhibit trends that are fully consistent
with those obtained from the 1D scheme in Fig. \ref{fig.2}. In particular,
both physical quantities provided by the three different methods show
essentially the same qualitative behavior as in the corresponding
1D calculations.

Based on these findings, it can be concluded that the stochastic SE
provides a reliable description of the dynamical processes occurring
in the open quantum system considered here. From a numerical perspective,
the stochastic SE offers several advantages over the Lindblad ME,
particularly in the context of modern computer architectures that
strongly benefit from parallelization. At the level of a single quantum
trajectory, the stochastic SE propagates a state vector rather than
a density matrix, thereby reducing the computational complexity from
$\mathcal{O}(N^{4})$ to $\mathcal{O}(N^{2})$, while simultaneously
lowering the associated memory requirements. Moreover, since individual
trajectories are statistically independent, stochastic SE simulations
can be parallelized straightforwardly, making the method highly attractive
even for moderately sized computational resources. Consequently, the
stochastic SE will serve as our reference method throughout the remainder
of this work. From now on, we restrict our analysis to the stochastic
SE and non-Hermitian TDSE solutions, and investigate the extent to
which the latter can reproduce the results of the reference approach.
Particular emphasis will be placed on identifying the range of validity
and the limitations of the non-Hermitian TDSE method in describing
lossy cavity-molecule dynamics.

In the next two figures, only 2D results are presented. In Fig. \ref{fig.4 },
the dynamics is initiated from two different initial conditions (
$|e,1\rangle$ and \textbar$g,2\rangle$), and we display again
the excited state molecular population and the mean photon number.
A comparison of the molecular population obtained with the two different
methods (Fig. \ref{fig.4 }A and Fig. \ref{fig.4 }C) reveals substantial
differences between the results. The effect of the incoherent decay
term $\gamma_{\textrm{c}}\hat{a}\hat{\rho}\hat{a}^{\dag}$, is clearly
visible. This term induces incoherent transitions $|\alpha,n+1\rangle\rightarrow|\alpha,n\rangle$,
which are not captured by the non-Hermitian TDSE. Although the molecular
populations obtained by the two approaches differ significantly, the
corresponding mean photon numbers are shown in Figs. \ref{fig.4 }B and
\ref{fig.4 }D remain remarkably similar. It is because, that the
photon loss itself is described in a comparable manner within both
formalisms. However, in the non-Hermitian TDSE approach, the wave
packet is continuously attenuated by the absorptive term $-\textrm{i}\frac{\gamma_{\textrm{c}}}{2}\hat{N}$.
As a consequence, the population associated with molecules that remain
in excited states after photon emission is progressively removed from
the simulation. The subsequent coherent exchange of excitation between
the molecule and the cavity mode can therefore no longer be followed
within the non-Hermitian TDSE framework.

In contrast, the stochastic SE explicitly accounts for quantum-jump
processes and thus retains the molecular population after photon loss
events. Consequently, it can correctly describe the continued excitation
exchange between the molecule and the cavity field following cavity
decay. This mechanism is responsible for the pronounced differences
observed in the molecular population dynamics, while only weakly affecting
the mean photon number.

In Fig. \ref{fig.5 }, the initial conditions are prepared by transferring
population from the $|g,0\rangle$ state of the system to the excited
polaritonic manifold using laser pulses with durations of $T=15,30\,and\,45\,\mathrm{fs}$.
As the pulse duration increases, the discrepancy between the stochastic
SE and the non-Hermitian TDSE approaches become more pronounced.
The overall behavior of the curves remains qualitatively similar to
that observed previously in Fig. \ref{fig.3} and Fig. \ref{fig.4 }.
However, the quantitative differences between the two methods increase
significantly with increasing pulse length.

As the laser pulse becomes longer, the non-Hermitian TDSE approach
gradually loses its ability to accurately reproduce the reference
results obtained from the stochastic SE. For the longest pulse considered
($T=45\,\mathrm{fs}$), the breakdown of the non-Hermitian TDSE description
becomes particularly pronounced, affecting both the excited state
population and the mean photon number. For the mean photon number,
a shorter time window is displayed because on longer timescales, no
additional relevant dynamical features emerge, as a substantial fraction
of the cavity photons has already been lost by that time. 

\begin{figure}
\includegraphics[width=0.5\textwidth]{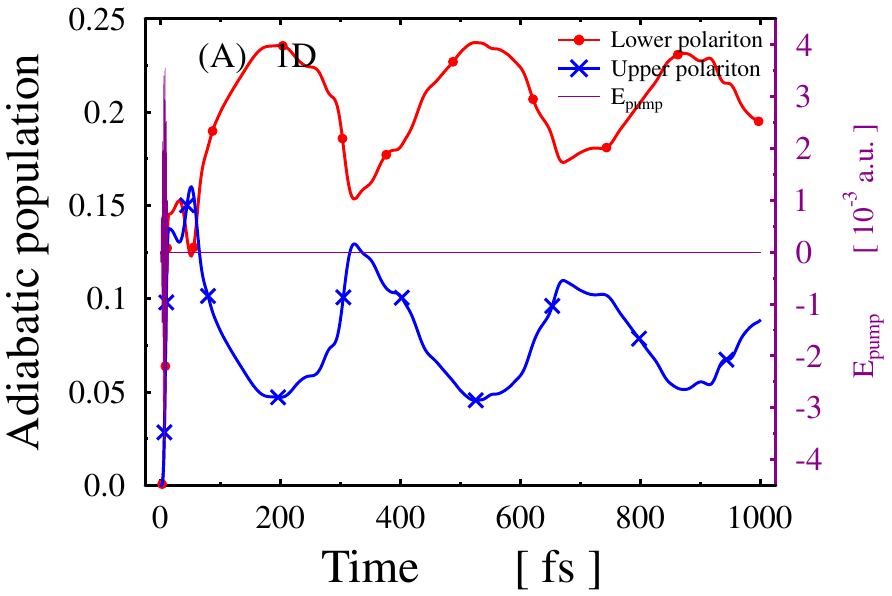}\includegraphics[width=0.5\textwidth]{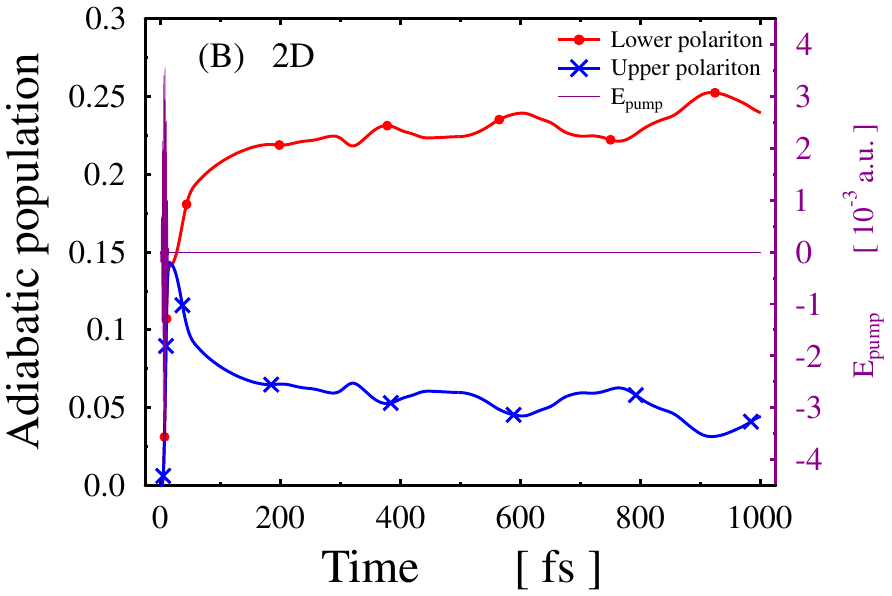}

\includegraphics[width=0.5\textwidth]{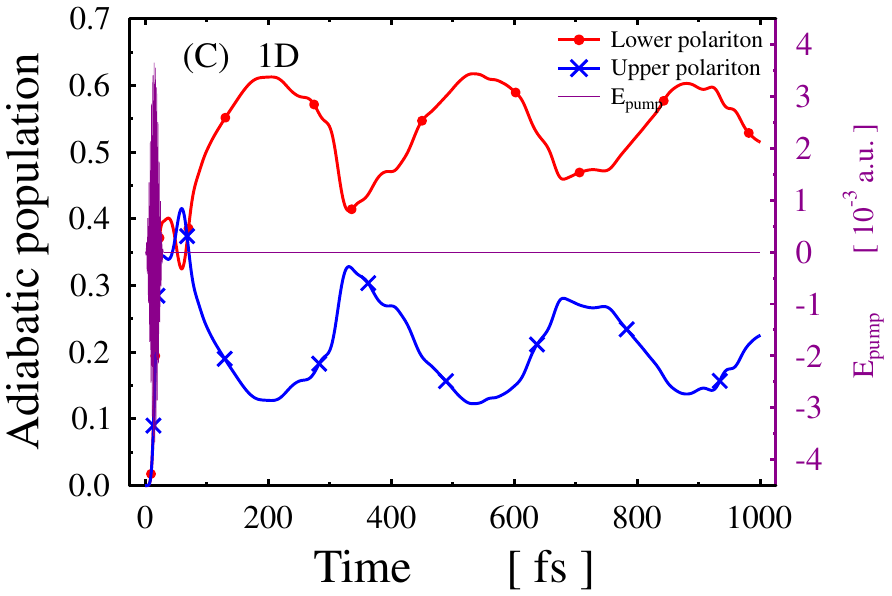}\includegraphics[width=0.5\textwidth]{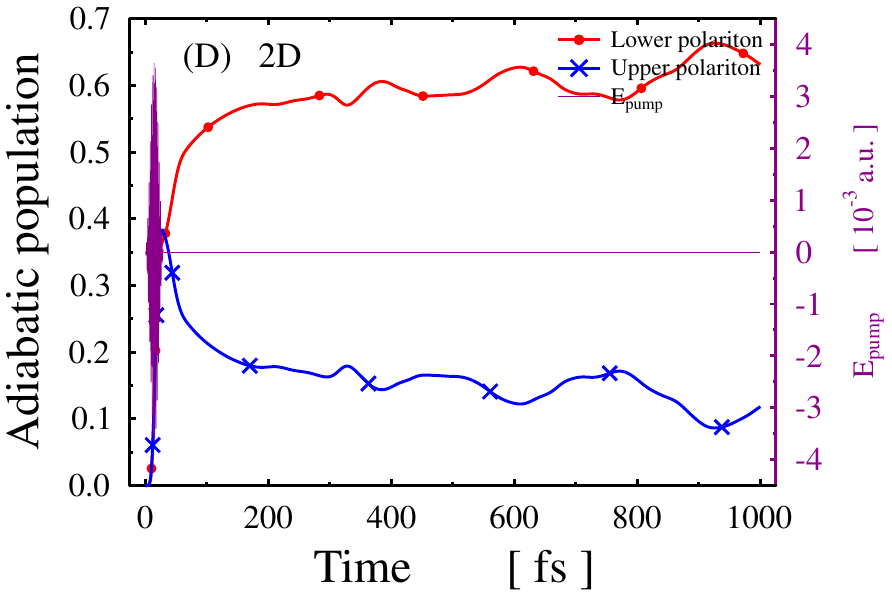}

\includegraphics[width=0.5\textwidth]{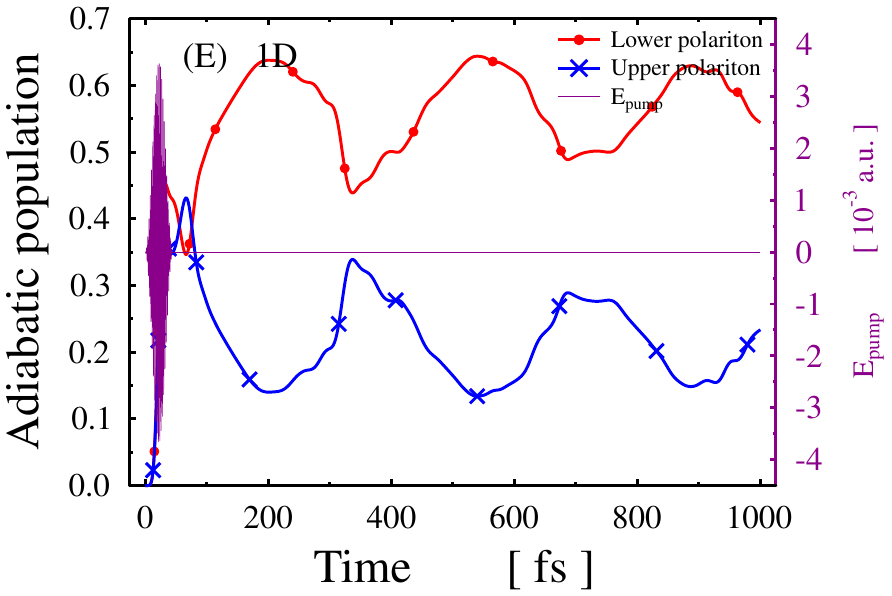}\includegraphics[width=0.5\textwidth]{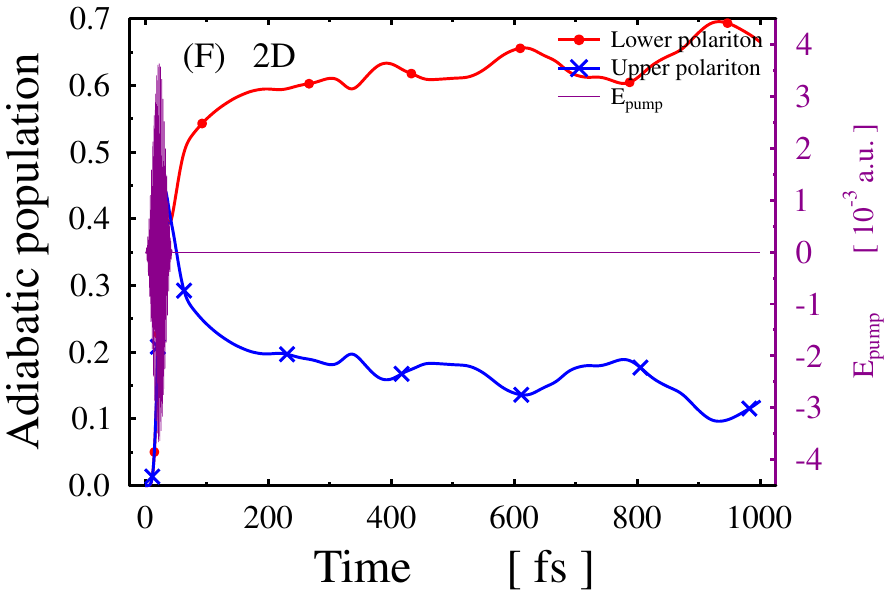}

\caption{\label{fig.6} Population dynamics of the 1LP and 1UP states in the
1D (Panels, A, C, and E) and 2D (Panels, B, D, and F) frameworks in the cavity
($g=5\times10^{-4}a.u.$ and $\omega_{cav}=1.968eV$) using the standard
TDSE method. The dynamics is initiated from the $|g,0\rangle$ state
pumping the population to the excited polaritonic manifold using a laser
pulses as ($I=4.74\times10^{11}w/cm^{2}$ and $\omega_{laser}=2eV$) with length of $T=15\,\mathrm{fs}$ (panels A and B), $T=30\,\mathrm{fs}$
(panels C and D) and $T=45\,\mathrm{fs}$ (panels E and F). }
\end{figure}

We now turn to the investigation of light-induced nonadiabatic molecular
dynamics. 
% On %% Krisztian
Fig. \ref{fig.6} we present the populations of the upper
(1UP) and lower (1LP) polaritonic surfaces obtained using the 1D and
2D schemes. It is known that polaritonic potential energy surfaces
are strictly valid only when the molecular axis is aligned with the
polarization direction of the cavity field. In contrast, when the
molecular axis is perpendicular to the cavity-field polarization,
the light-matter coupling vanishes and the diabatic representation
provides the appropriate description. For a rotating molecule, however,
the molecular orientation continuously evolves between these two limiting
cases. As a consequence, the diabatic and adiabatic (polaritonic)
surfaces are continuously transformed into one another as the molecular
orientation changes. The upper and lower polaritonic surfaces are
therefore not completely isolated. Instead, molecular rotation gives
rise to a LICI between them. At the LICI, the energy gap between the
adiabatic surfaces vanishes and the associated nonadiabatic couplings
become singular, enabling highly efficient population transfer between
the 1UP and 1LP states.

In the present simulations, the population is transferred from the $|g,0\rangle$
state of the system to the excited polaritonic manifold using a laser
pulses with durations of $T=15,30\,\textrm{and}\,45\,\mathrm{fs}$ and the
lossless dynamics are described within the standard Hermitian TDSE
framework. 

A striking difference is observed between the population dynamics
predicted by the 1D and 2D models. In the 1D approach, calculations
are performed for fixed molecular orientations, and rotational motion
is absent. Consequently, the molecular axis remains fixed with respect
to the cavity-field polarization throughout the propagation. Under
these conditions, the dynamics are characterized by oscillatory population
exchange between the 1UP and 1LP surfaces. In contrast, the 2D model
explicitly incorporates dynamical molecular rotation. As the molecule
continuously rotates during the dynamics, the wave packet repeatedly
encounters regions of strong nonadiabatic coupling associated with
the LICI. As a result, the population continuously leaks from the upper
polaritonic surface to the lower one. This behavior is clearly reflected
in the population curves: the population of the 1UP state decreases
steadily with time, while the population of the 1LP state exhibits
a corresponding increase. The observed population transfer demonstrates
the crucial role played by light-induced nonadiabatic effects and
highlights the limitations of reduced-dimensionality models that neglect
molecular rotation.

\begin{figure}
\includegraphics[width=0.5\textwidth]{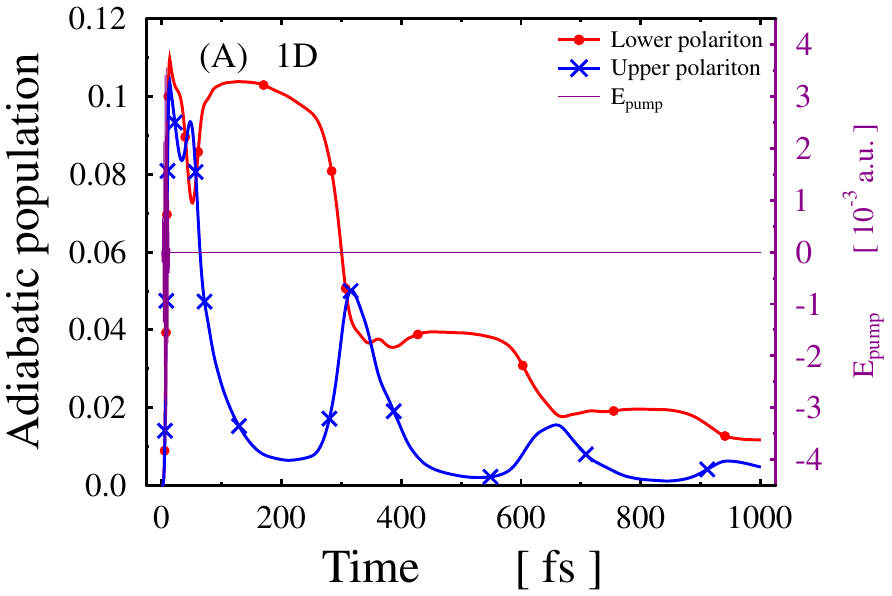}\includegraphics[width=0.5\textwidth]{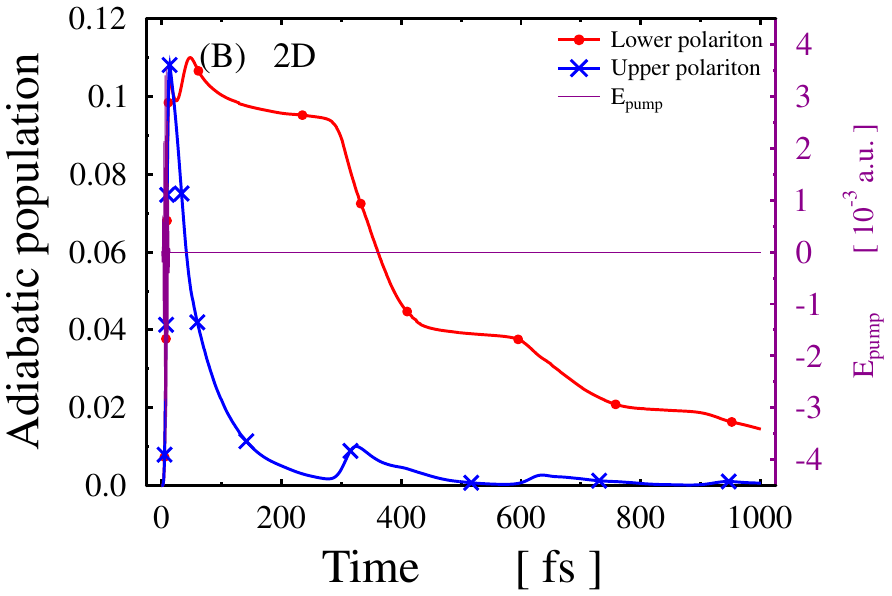}

\includegraphics[width=0.5\textwidth]{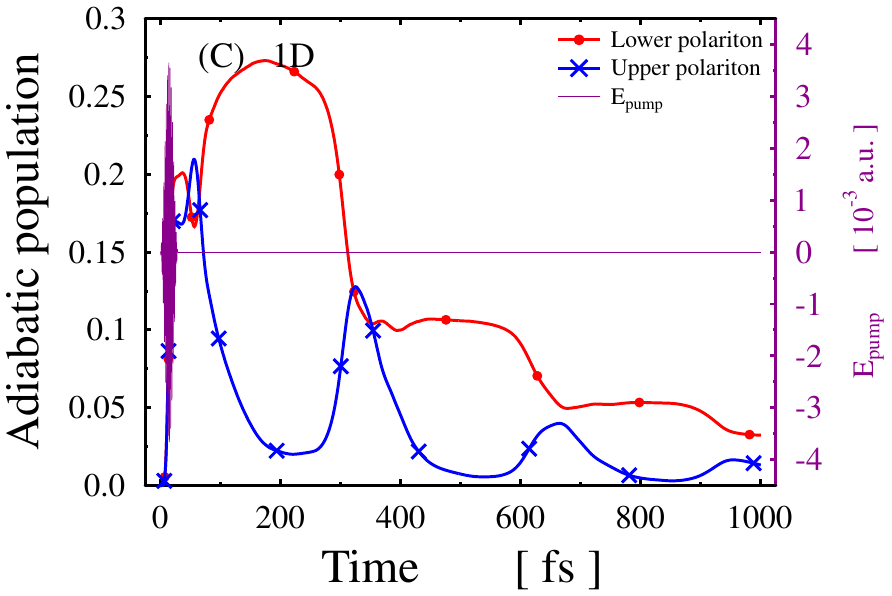}\includegraphics[width=0.5\textwidth]{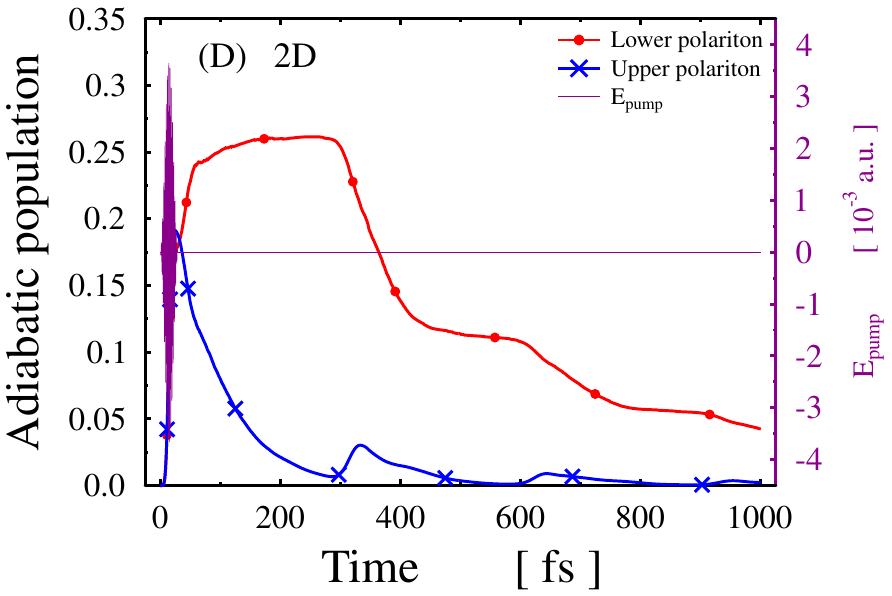}

\includegraphics[width=0.5\textwidth]{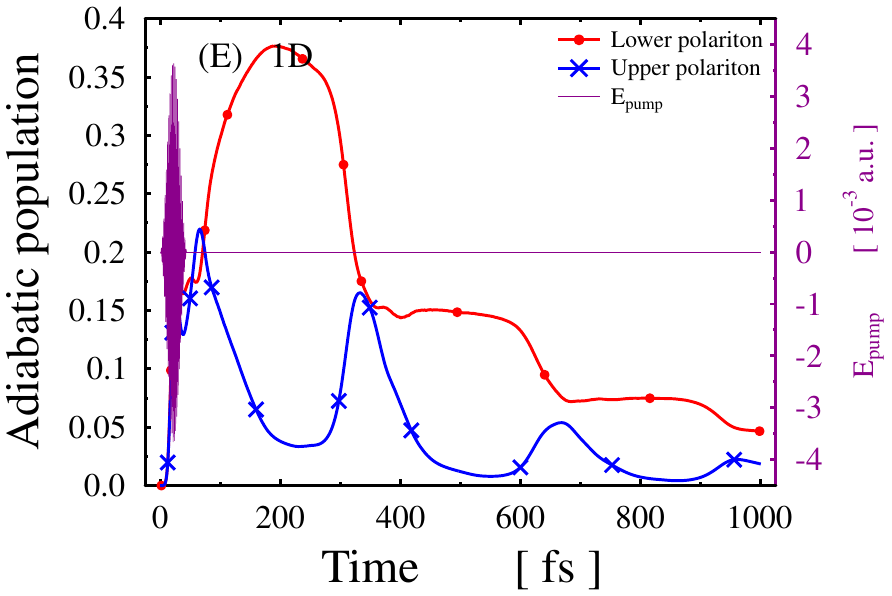}\includegraphics[width=0.5\textwidth]{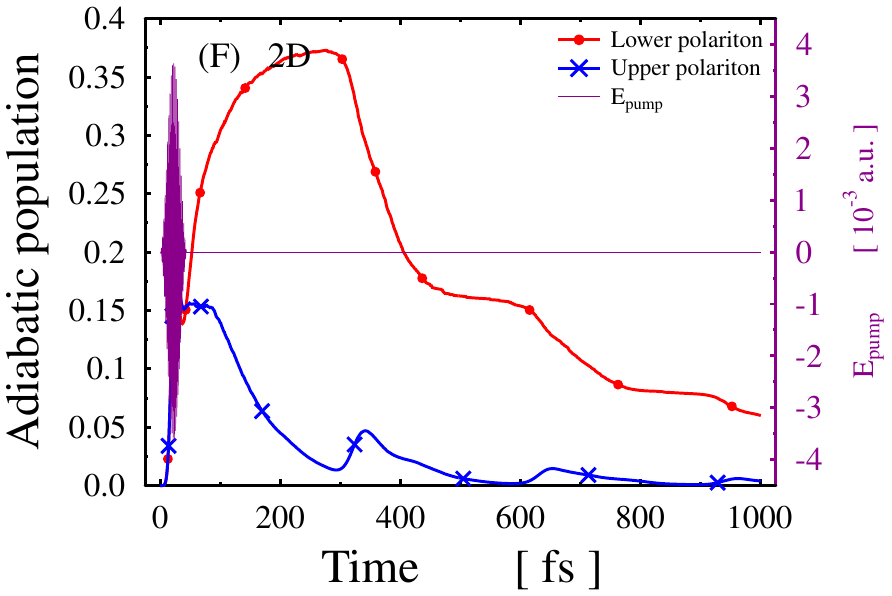}

\caption{\label{fig.7} Population dynamics of the 1LP and 1UP states in the
1D (Panels, A, C, and E) and 2D (Panels, B, D, and F) frameworks in the cavity
($g=5\times10^{-4}a.u.,$ $\gamma_{c}=4\times10^{-4}a.u.$ and $\omega_{cav}=1.968eV$)
using the stochastic SE method. The dynamics are initiated from the
$|g,0\rangle$ state, pumping the population to the excited polaritonic
manifold using laser pulses as ($I=4.74\times10^{11}w/cm^{2}$ and
$\omega_{laser}=2eV$ ) with length of $T=15\,\mathrm{fs}$ (panels
A and B), $T=30\,\mathrm{fs}$ (panels C and D) and $T=45\,\mathrm{fs}$
(panels E and F). }
\end{figure}

Fig. \ref{fig.7} displays the populations of the upper (1UP) and
lower (1LP) polaritonic surfaces obtained using the 1D and 2D models
including the cavity photon leakage. Population is transferred from
the $|g,0\rangle$ state of the system to the excited polaritonic
manifold using laser pulses with durations of $T=15,30\,and\,45\,\mathrm{fs}$,
but in this case, open system dynamics is treated and described by
the stochastic SE framework. 

It is seen that the cavity losses have a stronger influence on the
dynamics than the population oscillations between the polaritonic
surfaces in both models. Nevertheless, the latter effect remains clearly
visible. As discussed above, molecular rotation continuously drives
the wave packet through regions of nonadiabatic coupling, thereby
facilitating population transfer between the polaritonic surfaces.
Consequently, distinct peaks emerge in the population curves. These
features are particularly pronounced in the 1D model, especially for
the longer laser pulses. Their suppression in the 2D simulations indicates
that the population is transferred more efficiently from the upper polaritonic
surface to the lower one. This behavior is a direct consequence of
the additional nonadiabatic population-transfer pathway provided by
the LICI. 

\section*{IV. Conclusions }

In the present work, we investigated the dynamics of an open cavity-molecule
quantum system and compared three different theoretical approaches
for its description. We analyzed the time evolution of the excited
state populations and the mean photon number under a variety of physically
relevant conditions. Our results demonstrate that the Lindblad ME
and the stochastic SE yield very similar and mutually consistent predictions
across all investigated scenarios. In contrast, the non-Hermitian
TDSE was found to possess a more limited range of applicability and
may lead to significant deviations when incoherent cavity-loss processes
play an important role..

We further demonstrated that, for diatomic molecules, the 1D and 2D
descriptions differ not only in their treatment of rotational dynamics
but also in their ability to capture light-induced nonadiabatic phenomena.
It is well established that a proper description of nonadiabatic molecular
dynamics requires at least two independent nuclear degrees of freedom.
In the case of diatomic molecules, these are provided by the vibrational
and rotational coordinates. Using the standard TDSE in the lossless
case and the stochastic SE in the presence of cavity losses, we analyzed
the population dynamics on the 1UP and 1LP surfaces. The results revealed
substantial differences between the 1D and 2D descriptions, highlighting
the crucial impact of light-induced nonadiabatic phenomena on the
cavity- controlled molecular dynamics. The population curves exhibit
oscillatory behavior with peaks of varying amplitudes. Stronger peaks
in 1D imply that less population leaves the upper surface, whereas
in 2D the LICI provides a more efficient relaxation channel from the
upper to the lower polaritonic surface, reducing the prominence of
the oscillatory peaks.

{The limitations of the present model are mostly a matter of scope and controlled approximations. 
Cavity loss is modeled with a Markovian Lindblad description (and reproduced efficiently via stochastic SE trajectories), which is a standard and often well-justified choice, but may not capture all features of plasmonic nano-cavities where the environment can be spectrally structured and exhibit memory effects. Likewise, the molecular model is intentionally compact (e.g., a limited electronic-state manifold with reduced-dimensional nuclear dynamics), which is appropriate for method benchmarking and for highlighting effects like LICIs in 2D, but it naturally means that additional electronic channels or more detailed electronic-structure effects are not represented. Finally, the work relies on common simplifying Hamiltonian choices (e.g., rotating wave approximations and neglecting dipole self-energy), which are typically reasonable in the targeted parameter regime, yet could become more sensitive when exploring stronger coupling, different detunings, or ultrashort driving pulses.}

In a forthcoming work, we plan to investigate light-induced nonadiabatic
quantum dynamics in dissociative diatomic molecules coupled to lossy
optical cavities. Particular emphasis will be placed on the calculation
of experimentally measurable physical quantities, like angular distributions
and spectroscopic signatures of the photodissociation fragments.

\begin{acknowledgement}
The authors are indebted to NKFIH for funding (Grant No. K146096).
This paper was supported by the University of Debrecen Program for
Scientific Publication. The ELI ALPS project (GINOP-2.3.6-15-2015-
00001) is supported by the European Union and co-financed by the European
Regional Development Fund. The work of P.B. and K.V. was supported by
the National Science Foundation (NSF) under Grant No. DMR-2217759 and 
Grant No. IRES-2245029.
\end{acknowledgement}

\section*{Data availability} 
The data that supports the findings of this study are available within the article.

\section*{Conflicts of interest}
The authors have no conflicts to disclose.

% Feist,  Stephan,  Lenz

\bibliography{cavity_dse,Methods}

\end{document}